\documentclass[aps,superscriptaddress,showpacs]{revtex4}
\usepackage{amsmath}
\usepackage{amssymb}
\usepackage{graphicx}
\usepackage{ulem}

\begin{document}
\title{Resonant Hawking radiation in Bose-Einstein condensates}
\author{I. Zapata}
\affiliation{Departamento de F\'{\i}sica de Materiales, Universidad Complutense de Madrid,
E-28040 Madrid, Spain}
\author{M. Albert}
\affiliation{D\'epartement de Physique Th\'eorique, Universit\'e de Gen\`eve,
CH-1211 Gen\`eve, Switzerland}
\author{R. Parentani}
\affiliation{Laboratoire de Physique Th\'eorique, 
CNRS UMR 8627, B\^at. 210, Universit\'e Paris-Sud 11,
91405 Orsay Cedex, France}
\author{F. Sols}
\affiliation{Departamento de F\'{\i}sica de Materiales, Universidad Complutense de Madrid,
E-28040 Madrid, Spain}

\pacs{03.75.Kk Dynamic properties of condensates; collective and hydrodynamic excitations, superfluid flow  04.62.+v Quantum fields in curved spacetime  04.70.Dy Quantum aspects of black holes, evaporation, thermodynamics}
\date{\today}

\begin{abstract}
We study double-barrier interfaces separating regions of asymptotically subsonic and supersonic flow of Bose condensed atoms.
These setups contain at least one black hole sonic horizon
from which the analog of Hawking radiation should
be generated and emitted against the flow in the subsonic region. Multiple coherent scattering by the double-barrier structure strongly modulates the transmission probability of phonons, rendering it very sensitive to their frequency. As a result, resonant tunneling occurs with high probability within a few narrow frequency intervals. This gives rise to highly non-thermal spectra with sharp peaks. We find that these peaks are mostly associated to decaying resonances and only occasionally to dynamical instabilities.
Even at achievable nonzero temperatures, the radiation peaks can be dominated by the spontaneous emission, i.e. enhanced zero-point fluctuations,
and not, as often in analog models, by stimulated emission.
\end{abstract}

\volumeyear{2010}
\volumenumber{number}
\issuenumber{number}
\eid{identifier}
\startpage{1}
\endpage{ }
\maketitle

\section{Introduction}

 Hawking radiation is one of the most intriguing yet unobserved predictions of modern physics.
It is believed to be generated near the horizon surrounding a black hole and to be responsible for its ultimate decay
\cite{hawking1974,Unruh1976}. As noted by Unruh \cite{Unruh1981}, a phenomenon analog to Hawking radiation
should occur near a sonic horizon in a moving fluid,
i.e. near the interface separating regions of subsonic and supersonic current.
He argued that a thermal flux of phonons
should be spontaneously generated from the sonic horizon towards the subsonic region.
The effect finds its origin in the impossibility of defining a global quasiparticle vacuum which suits both incoming and outgoing states.
More recently, it has been proposed that flowing condensates of bosonic atoms could provide interesting
analogs of black hole physics \cite{Garay2000,Laughlin2003} and in particular of their Hawking radiation \cite{balbinot2008,carusotto2008,macher2009,finazzi2010,coutant2010}. The interface between subsonic and supersonic regions has also been shown to provide a scenario for the bosonic analog of Andreev reflection \cite{zapata2009}.

Except for black hole lasers (discussed in Refs. \cite{finazzi2010,coutant2010,leonhardt2008,corley1999}),
most proposals predict a spectrum (phonon current distribution per unit frequency) with a single peak at frequency $\omega=0$ falling as $1/\omega$ for small $\omega>0$, where $\hbar\omega$ is the quasiparticle  energy measured with respect to the condensate chemical potential.
This means that the low-frequency peak is essentially thermal in character.
However, because of dispersive effects,
the effective temperature characterizing that zero-point radiation
is no longer universal. In a Bose-Einstein condensate, the group velocity of phonons increases at high frequency, unlike in the original model considered by Unruh \cite{Unruh1995}. As a result, this ``superluminal" transport dilutes the sonic horizon into a spatial interval of finite size \cite{finazzi2010,finazzi2011}. The blue-shifting effect accompanying with the sonic horizon implies that the dispersive properties of phonons are always involved. Nevertheless, when the dispersive length scale is much smaller than the horizon curvature scale, the temperature is determined by the local properties of condensate flow near the horizon (the gradient of the flow~\cite{Unruh1981})
in strict analogy with the standard Hawking radiation which is fixed by the surface gravity of the black hole.
This regime is found when the gradient is much smaller than one in units of the healing length~\cite{macher2009}.
Instead, when it is higher than one, the dispersion effects dominate and as a result
the effective temperature is fixed by the  healing length and the
jump of the velocities across the sonic horizon~\cite{recati2009}.
In any case, the effective temperature will be smaller than the chemical potential \cite{lahav2010}.
As a consequence, a direct attempt to measure this radiation profile appears extremely difficult.
An alternative proposal to indirectly measure Hawking radiation relies on the squeezed character of the
state \cite{balbinot2008,carusotto2008} which could be observed at currently attainable temperatures by counting atoms on both sides of the horizon at coinciding times. However it should be pointed out that the main contribution to these correlations are due to the stimulated
amplification of pre-existing phonons \cite{carusotto2008,macher2009},
and not to the spontaneous amplification of vacuum fluctuations.

Here we study a new method which specifically aims at detecting the spontaneous contribution to
Hawking radiation. This approach relies on the strong frequency-dependence of resonant tunneling through a double barrier structure. Such a sonic black-hole analog behaves as a Fabry-Perot resonator for quasiparticles, with the peculiar feature that quasiparticles propagate linearly against a condensate background which is itself governed by a nonlinear equation. The Hawking emitted phonon spectrum shows peaks at frequencies different from zero. We find that, at currently achievable temperatures, thermal noise could be weak enough not to blur the characteristics of this resonant radiation, and a time-of-flight experiment could allow for its detection. Our proposal is quite similar to the black-hole laser setup \cite{finazzi2010,coutant2010,leonhardt2008,corley1999} in that
sharp peaks are found in both cases. However, although dynamical instabilities may appear occasionally, they are not a necessary feature of these type of setups (see Section \ref{QNM.sect}). We propose here to focus on situations that are dynamically stable, i.e. where all peaks are due to resonances. The dynamical stability of the flow is likely to be a valuable asset in actual experiments. A systematic study of resonances and instabilities is currently under investigation.

 This paper is arranged as follows: In section II we present a mean-field study of the considered setup, discuss the main results and some general features. Section III is devoted to formulate the scattering problem of Bogoliubov quasiparticles propagating against the condensate background and to identify the essential features of Hawking radiation. In section IV we present and discuss the numerical results obtained for the Hawking radiation spectrum emitted from a double delta-barrier interface separating a subsonic and a supersonic region. Section V deals with the general distinction between quasinormal modes (or resonances) and dynamical instabilities, both of which are candidates to lie behind the sharp peaks in the radiation spectrum. Finally, Section VI is devoted to a summary and discussion of the main results. The main text is complemented by two appendices. Appendix A presents the analytical calculation of the mean-field model. Appendix B deals with the analytical resolution of the quasiparticle eigenvalue problem in the inhomogeneous region on the subsonic and supersonic sides near the double-barrier interface.

\section{Formulation of the model. Condensate wave function.}

\begin{figure}[htb!]
\includegraphics[width=\columnwidth]{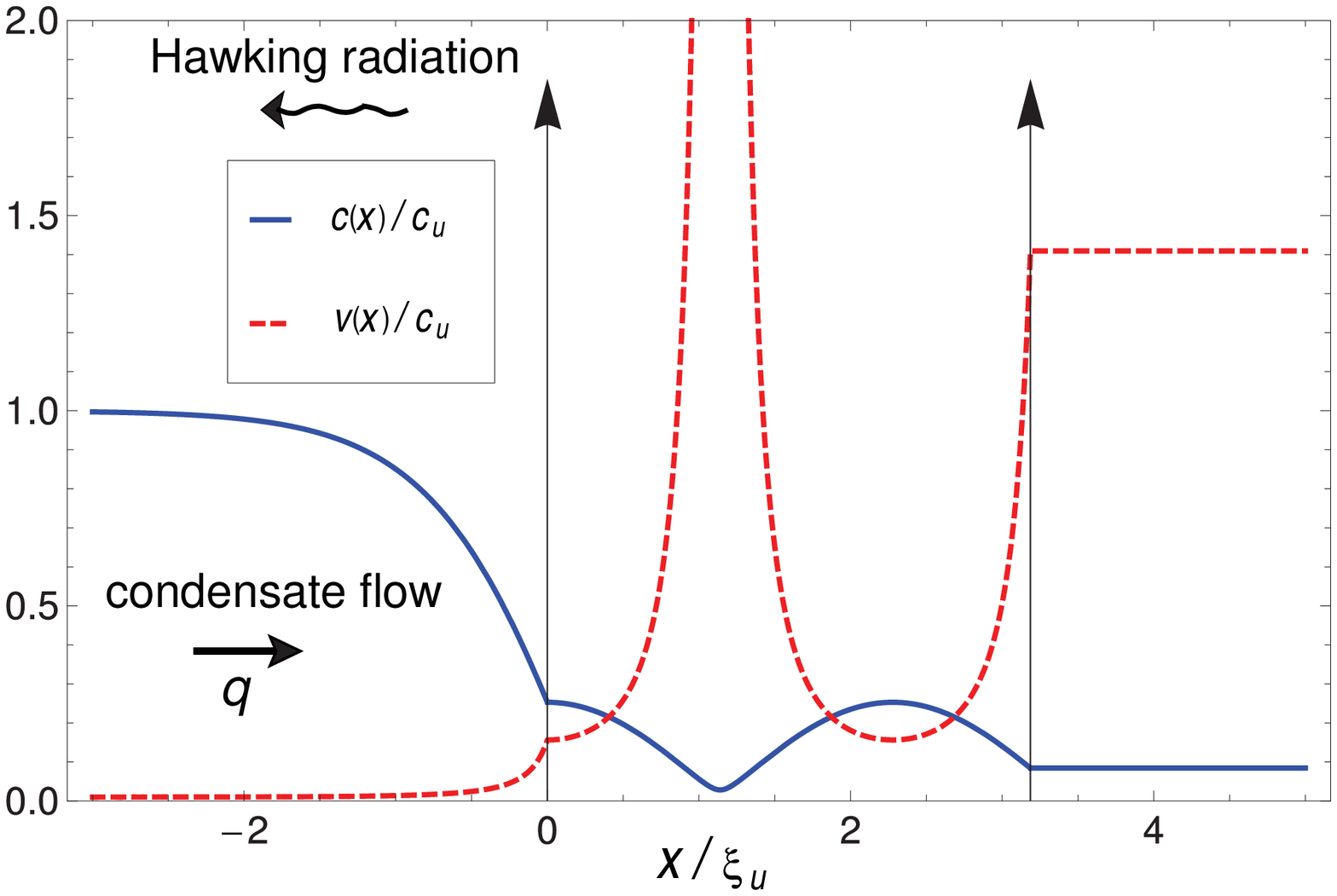}
\caption{A typical black-hole structure discussed in this paper. The two vertical arrows represent the delta barriers. The profiles of the speed of sound [proportional to $\sqrt{\rho(x)}$, where $\rho(x)$ is the local condensate density] and the speed of flow [proportional to $1/ \rho(x)$] are shown. Condensate flows to the right, while Hawking radiation is emitted into the left (subsonic) region. In this particular case, three event horizons exist since the two curves intersect at three points. The flow regime is such that the condensate density exhibits a depression between the barriers. }
\label{figDDSamplePlot}
\end{figure}

We study an atom transport setup which is schematically depicted in Fig. \ref{figDDSamplePlot}. A quasi-1D bosonic condensate, which occupies the left region $x<0$, is allowed to leak to the right though two identical delta potential barriers. The leftmost barrier is conventionally placed at $x=0$ separated by a distance $d$ from the second barrier. We will see that this double-barrier setup behaves as a resonant structure. For convenience, we neglect quantum fluctuations of the condensate stemming from its one-dimensional character \cite{jackson1998,leboeuf2001}.

We may decompose the Heisenberg second-quantized field operator
\begin{equation}\label{totalfield.eq}
\widehat{\Psi}(x,t)=e^{-i \mu t/\hbar}\Psi_0(x)+\delta\widehat{ \Psi}(x,t)
\end{equation}
into a stationary condensate wave-function $e^{-i \mu t/\hbar}\Psi_0(x)$, with $\mu$ the chemical potential, and its fluctuations $\delta\widehat{\Psi}(x,t)$. In this section we focus on the condensate behavior. At low temperatures and densities, the mean-field equation which governs the stationary flow of a Bose-Einstein condensate is the time-independent Gross-Pitaevskii (GP) equation for the condensate wave function:
\begin{equation}\label{GPDim.eq}
\left[-\frac{\hbar^2}{2m} \frac{\partial^2}{\partial x^2} - \mu + V_{\rm ext}(x)+g_{\rm 1D}|\Psi_0(x)|^2\right] \Psi_0(x)=0 \; .
\end{equation}

We wish to study the effect of a potential consisting of two delta barriers of equal strength, although for comparison we will occasionally consider the single barrier case (see Ref. \cite{pavloffTBP}). Thus we will assume an external potential $V_{\rm ext}(x)$ which takes one of the two following forms:

\begin{equation}\label{potential.eq}
V_{\rm ext}(x)=\left\{
           \begin{array}{c}
             V_{1}(x)\equiv \hbar c_u z\delta(x) \\
             V_{2}(x)\equiv  \hbar c_u z\left[\delta(x)+\delta(x-d)\right]
           \end{array}\right. \; ,
\end{equation}
where $c_u=\sqrt{g_u n_u / m}$, with $n_u \equiv \lim_{x \rightarrow -\infty} |\Psi_0(x)|^2$, is the speed of sound on the subsonic, upstream ($x<0$) side, and $z$ the dimensionless strength of each barrier. The healing length on the asymptotic upstream side is $\xi_u=\hbar/mc_u$.

In both the single- and double-barrier case, solutions can be found in which the condensate velocity is supersonic on the right of the second barrier, and subsonic from  $-\infty$ to some point in the vicinity of the leftmost ($x=0$) barrier (see Appendix \ref{MFT.app} for a general discussion of these solutions). Those flow profiles must have one or more horizons, defined as points where the local condensate velocity equals the local speed of sound. This subsonic-supersonic scenario is the same which has been shown to display the bosonic analog of Andreev reflection, where the supersonic side plays the role of the normal fluid \cite{zapata2009}. Thus one may generally speak of Andreev-Hawking processes when dealing with scattering events undergone by elementary excitations of condensate flow through a subsonic-supersonic interface.

Such sonic analogs relying on condensate flow have a superluminal dispersion relation at higher frequencies and their horizons do not imply a strict causal disconnection among regions, but they are sufficient to produce Hawking radiation analogs (see however Ref. \cite{barcelo2006a} for a discussion of a scenario where horizons would not be needed). For the case of a single delta barrier the only horizon lies on the near left of the barrier. The double barrier case is richer: there appears one horizon in the vicinity of the $x=0$ barrier (which may lie on either side) and possibly one or more horizon pairs. The leftmost horizon, and the only one in the single delta barrier case, can be viewed as a black hole (BH) analog, because there the flow goes from the subsonic to the supersonic side. The possible additional pairs of horizons appearing in a double barrier structure are the analogs of white-hole/black-hole pairs, where a white hole (WH) is the time-reversed version of a BH.
White hole analogs seem to be extremely difficult to generate experimentally. There is a debate in the literature on whether white holes are stable at all, a question to which
linear stability analysis has not yet provided an answer \cite{macher2009,mayoral2011}.

To clearly identify what constitutes Hawking radiation, the experiment should ideally be done in such a way that the condensate wave-function is stationary and asymptotically flat: $\rho'(\pm\infty)=0$, where $\rho(x)\equiv|\Psi_0(x)|^2$ \cite{birrell1982}. In Appendix \ref{MFT.app} we show that, with those homogenous boundary conditions, there is one or more mean-field solutions for two delta potential barriers,  and just one for a single delta barrier. A sketch of a typical density for two delta barriers is shown in Fig. \ref{figDDSamplePlot}.

\begin{figure}[htb!]
\includegraphics[width=\columnwidth]{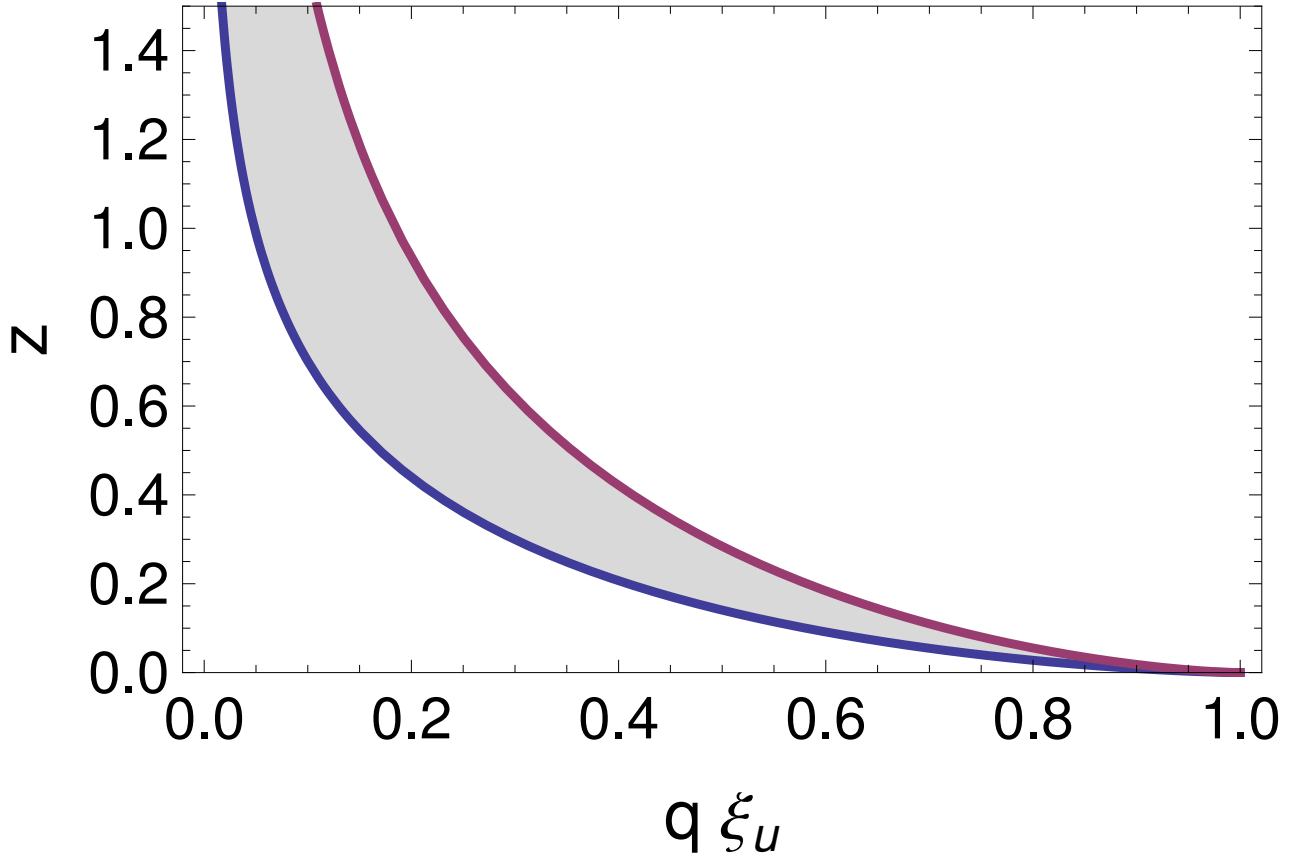}
\caption{For a structure of two delta barriers of identical strength, the region of $z,q$ for which solutions exist is shown in grey color. The minimum and maximum delta barrier strengths $z$ are plotted as a function of the current $q$ (which in these units coincides with the upstream condensate momentum). Red curve shows the maximum $z$ for a double delta barrier system, which is the same as the unique $z$ value for a single delta barrier of the same strength [see Eq. (\ref{1deltaMatch.eq})]. The blue line is the minimum $z$ value allowing for a solution in the double barrier case [see Eq. (\ref{match2DeltasRho.eq}) and the ensuing paragraph].}
\label{figDeltaStrengthPlot}
\end{figure}

In Fig. \ref{figDeltaStrengthPlot} we plot the region of the $(z,q)$ plane for which stationary solutions with homogeneous boundary conditions exist to the double-barrier problem. We recall that $z$ is the dimensionless parameter characterizing the strength of the two identical delta barriers, while $q$ is the condensate momentum on the subsonic (upstream) side. We note that, for a given value of $q$, there is a minimum and a maximum barrier strength between which a solution is guaranteed to exist for that particular value of $q$. This contrasts with the behavior of the single barrier case, for which only one solution exists for a given value of $q$. Interestingly, the resulting line $z(q)$ line for the single-barrier case coincides with the upper boundary of the shaded region in Fig.  \ref{figDeltaStrengthPlot}.

\begin{figure}[htb!]
\includegraphics[width=\columnwidth]{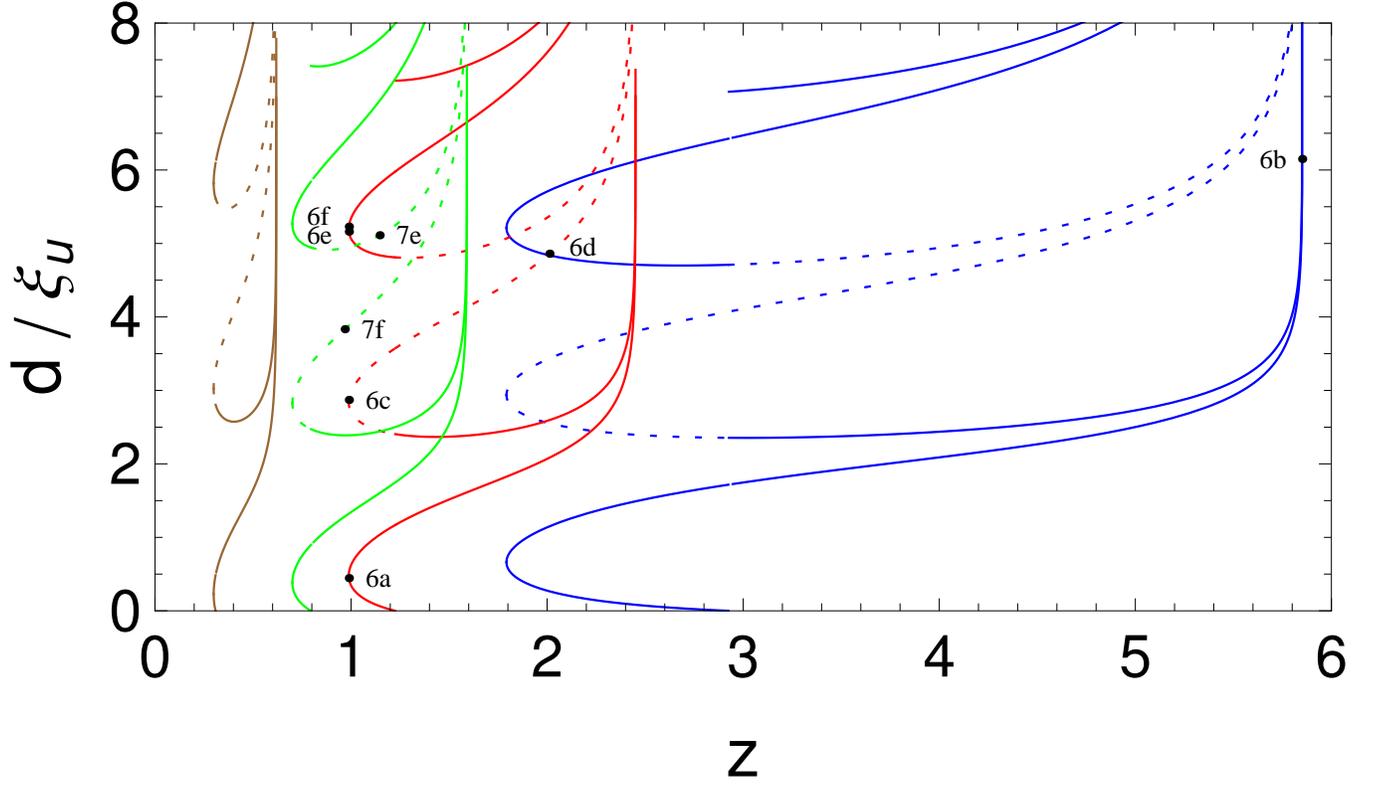}
\caption{Strength $z$ of the double delta barrier as function of the inter-barrier distance $d$. The plotted lines ranging from right to left (with colors blue, red, green, brown) correspond to $q\xi_u=0.01, 0.05, 0.1, 0.3$. Solid lines stand for solutions whose density profile undergoes zero or two oscillations between barriers, while the dotted lines correspond to solutions with one full oscillation. The multiplicity of the values of $z$ as a function of $d$ reflects the possibility of nonlinear oscillations between the barriers. The dots, with their respective labels, correspond to the various configurations shown in Fig. \ref{grHawkingPlot} (with velocity profiles given in Fig. \ref{grDdPlot}) and Fig. \ref{grTOFPlot}.}
\label{figPhaseDiagplot}
\end{figure}

Figure \ref{figPhaseDiagplot} shows a representative sample of solutions in the $(d,z)$ plane, where $d$ is the distance between barriers. Each curve represents a value of the condensate momentum $q$. Inspection of Fig. \ref{figPhaseDiagplot} reveals that, for a given $q$, there is a finite interval of allowed $z$ values. That range has been shown in Fig. \ref{figDeltaStrengthPlot}. Figure \ref{figPhaseDiagplot} reveals that, for given $z$ and $q$, a multiplicity of $d$ solutions exist. The two lowest $d$ values can be characterized by two different values of $\rho(0)$, which we call $\rho_{\rm max}$ and $\rho_{\rm min}$. The upper $d$ solutions come also in pairs and are regularly separated by a distance difference equal to the period of the nonlinear oscillations between the two barriers, which is the same for both $\rho_{\rm max}$ and $\rho_{\rm min}$.

An interesting feature is that, at small barrier separations, a single $q$ solution exists for a given $d$ and $z$. For higher $d$ (slightly above 2 for the sample of curves shown in Fig. \ref{figPhaseDiagplot}), two $q$ solutions exist for given $d$ and $z$. For still higher $d$, three $q$ solutions exist for given $d$ and $z$, and so forth. Since the barrier separation $d$ and the barrier strength $z$ are expected to be experimentally adjustable parameters, the clear trend is that, the larger the distance, the higher the number of allowed stationary solutions, a fact which could translate itself into instabilities.

\section{Bogoliubov analysis}\label{BdG.sect}

A general and detailed introduction to the Hawking radiation physics in Bose-Einstein condensates, within a Bogoliubov-deGennes description can be found in Refs. \cite{recati2009,macher2009,leonhardt2003,leonhardt2003A}. In this section we mainly introduce notation while refer the reader to those works for a more complete presentation.

The quantum fluctuation part introduced in Eq. (\ref{totalfield.eq}), $\delta\widehat{ \Psi}(x,t)$, can be subject to a canonical transformation resulting in the expansion
\begin{equation}\label{BogDecomp.eq}
\delta\widehat{ \Psi}(x,t)= e^{-i \mu t/\hbar} \sum_i \left[ u_i(x)e^{-i\omega_i t} \hat{\gamma}_i+v_i^*(x)e^{i \omega_i t} \hat{\gamma}^\dag_i \right] \; ,
\end{equation}
where $u_i(x)$ and  $v_i^*(x)$ are the components of the wave function of the bosonic quasiparticle created by $\hat{\gamma}^\dag_i$. These $u,v$ components satisfy the bosonic Bogoliubov-deGennes (BdG) equations:
\begin{eqnarray}\label{BdGDim.eq}
\hbar \omega \left[
                               \begin{array}{c}
                                 u(x) \\
                                 v(x) \\
                               \end{array}
                             \right]
                             &=&
                             \left[
                               \begin{array}{cc}
                               \hat{H}& g_{\rm 1D} \Psi_0(x)^2 \\
                               -g_{\rm 1D} \Psi_0^*(x)^2 & -\hat{H} \\
                               \end{array}
                             \right]
                             \left[
                               \begin{array}{c}
                                 u(x) \\
                                 v(x) \\
                               \end{array}
                             \right]\\
\nonumber
\hat{H}&\equiv&-\frac{\hbar^2}{2m} \frac{\partial^2}{\partial x^2} - \mu + V_{\rm ext}(x)+2 g_{\rm 1D}|\Psi_0(x)|^2 \; ,
\end{eqnarray}
while the $\hat{\gamma},\hat{\gamma}^\dag$ operators satisfy, for real $\omega$,
\begin{equation}\label{commutator.eq}
\left[ \hat{\gamma}_i, \hat{\gamma}^\dag_j \right] = \int dx \left[ u_i^*(x) u_j(x) - v_i^*(x) v_j(x) \right]=\nu_i \delta_{ij} \; ,
\end{equation}
where the normalization $\nu_i$ can be set to $\pm 1$.

In Eq. (\ref{BdGDim.eq}) $g_{\rm 1D}=2\hbar^2 a_0 /(m a_\bot^2)$ is the effective 1D coupling constant, where $a_0$ is the s-wave scattering length and $a_\bot$ the transverse confinement harmonic oscillator length, and $V_{\rm ext}(x)$ is an externally imposed potential. We note that the sum over $i$ in the first equation may be interpreted as including both $\omega_i\gtrless0$, with $\nu_i>0$, or both $\nu_i\gtrless0$, with $\omega_i>0$.

The role of the horizon is appreciated when doing a WKB-type approximation for the excitations (see Refs. \cite{leonhardt2003,leonhardt2003A}). Close to the horizon are the turning points of the low energy classical trajectories where the WKB-solution is not appropriate. By matching the solutions on both sides and assuming some general scale separation (see also Ref. \cite{unruh2005}), one can show that in the case of just one horizon the profile of emitted Hawking radiation approaches $1/\omega$ at low frequencies. This important result guarantees the (approximately) thermal radiation profile. Thus at low frequencies, and in the absence of other scattering obstacles, the zero-point radiation has a $1/\omega$ behavior which makes it in principle difficult to distinguish from truly thermal behavior.

Due to the presence of delta barriers in our chosen configuration, WKB-type approximations cannot be used uncritically. However, thanks to a theorem on dark-soliton perturbation theory (see Ref. \cite{chen1998}) we know that we have at our disposal a complete set of solutions on the left of the $x=0$ barrier. In the flat supersonic region ($x>d$) the solutions are even simpler to work out. We refer the reader to Appendix \ref{BdG.app} for both cases, $x<0$ and $x>d$. The only non-analytically solvable problem lies between barriers ($0<x<d$), but this is a finite region where numerical integration of the BdG equations (\ref{BdGDim.eq}) requires only a moderate computational effort.

\begin{figure}[htb!]
\includegraphics[width=\columnwidth]{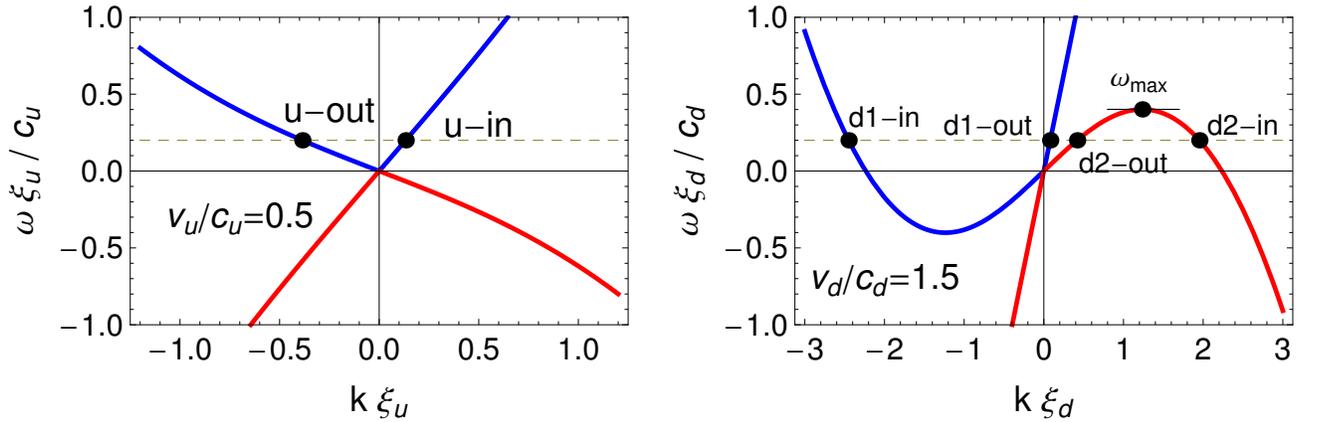}
\caption{Dispersion relation [see Eq. (\ref{BogDR.eq})] on the subsonic (left) and supersonic (right) sides. The blue/red branches correspond to positive/negative normalization as defined in Eq. (\ref{commutator.eq}).}
\label{figDispRelation}
\end{figure}

The next step is to identify the relevant scattering states. This has been done for Bose-Einstein condensates in studies  of BH analogs \cite{recati2009,macher2009} and of bosonic Andreev reflection \cite{zapata2009}. In this article we focus on the scattering states and their connection to quasi-normal modes (or resonances) and dynamical instabilities. A systematic study of complex-energy eigenmodes is left for a future work \cite{zapataTBP}. In the asymptotic regions, propagating modes obey the Bogoliubov dispersion relation. Following Ref. \cite{recati2009} we label the asymptotic regions with indeces $u$, for upstream ($x\rightarrow-\infty$), and $d$ for downstream ($x\rightarrow\infty$). The upstream dispersion relation is
\begin{equation}\label{BogDR.eq}
\omega_u(k)=v_{u} k \pm c_{u}|k|\sqrt{1+(k\xi_{u})^2/4} \; ,
\end{equation}
where $v_u$ is the upstream flow velocity, and similarly for downstream $\omega_d(k)$.
A graph of this relation is shown in Fig. \ref{figDispRelation} for both the subsonic and supersonic side. The branches shown in blue/red correspond to the $+/-$ of Eq. (\ref{BogDR.eq}) and can be shown to lead to positive/negative normalization. Modes are named after the sign of their group velocity (in/out according to whether they approach/leave the scattering structure), location (u/d) and, in the supersonic case, 1-2 stands for modes with normal/anomalous (i.e. positive/negative) normalization. Importantly, the anomalous d2 modes exist only for frequencies $\omega < \omega_{\rm max}$.

Here we are mainly interested in the scattering state with frequency $\omega$ characterized by the incoming channel d2-in. Its wave function reads

\begin{equation}\label{d2-in.eq}
{u_{\rm{d2-in, \omega}}(x)\brack v_{\rm{d2-in, \omega}}(x)}=\left\{
\begin{array}
[c]{ll}
\psi_u(x), & x\rightarrow-\infty\\
\psi_d(x), & x\rightarrow\infty
\end{array}
\right.
\end{equation}

where

\begin{equation}\label{psid.eq}
\begin{split}
  \psi_d(x)&={u_{d2}(k_{\rm{d2-in}}) e^{i q_d x} \brack v_{d2}(k_{\rm{d2-in}}) e^{-i q_d x}} \frac{e^{i k_{\rm{d2-in}} x}}{\sqrt{2\pi |w_{d2}(k_{\rm{d2-in}})|}}
\\ &+S_{\rm{d1d2}}(\omega) {u_{d1}(k_{\rm{d1-out}}) e^{i q_d x}\brack v_{d1}(k_{\rm{d1-out}}) e^{-i q_d x}} \frac{e^{i k_{\rm{d1-out}} x}}{\sqrt{2\pi |w_{d1}(k_{\rm{d1-out}})|}}\\
&+S_{\rm{d2d2}}(\omega){u_{d2}(k_{\rm{d2-out}}) e^{i q_d x}\brack v_{d2}(k_{\rm{d2-out}}) e^{-i q_d x}} \frac{e^{i k_{\rm{d2-out}} x}}{\sqrt{2\pi |w_{d2}(k_{\rm{d2-out}})|}},
\end{split}
\end{equation}

\begin{equation}\label{psiu.eq}
  \psi_u(x)=S_{\rm{ud2}}(\omega){u_u(k_{\rm{u-out}}) e^{i q_u x}\brack v_u(k_{\rm{u-out}}) e^{-i q_u x}} \frac{e^{i k_{\rm{u-out}} x}}{\sqrt{2\pi |w_u(k_{\rm{u-out}})|}}
\end{equation}
where $q_u,q_d$ are the up- and down-stream condensate momenta, $w=d\omega/dk$ denote the relevant group velocities, and the $S$-matrix elements are shown. In Eqs. (\ref{d2-in.eq})-(\ref{psiu.eq}) $u_\alpha(k)$ and $v_\alpha(k)$ are the spinor components for a quasiparticle of the uniform Bose gas, in channel $\alpha$ at momentum $k$. A similar decomposition can be made for the other in-modes. We note that the $k$'s of the previous equation are solutions of Eq. (\ref{BogDR.eq}) for a given $\omega$ lying in the interval $0<\omega<\omega_{\rm max}$, each solution defining a particular mode. Not shown in the previous equation but necessary for the matching, an evanescent solution exists on the subsonic side. The same holds on the supersonic side for $\omega>\omega_{\rm max}$.
To avoid double counting one can either choose all the positive normalization modes (for both in and out), as is done in Ref. \cite{zapata2009}, and then one has to deal with negative frequency modes, or as is usual in this Hawking context, choose only positive frequency modes and then interchange $\hat{\gamma}_i \leftrightarrow -\hat{\gamma}^\dag_i$ for the negative normalization ones \cite{footnote1}. In this work we adopt the convention of $\omega_i>0$ and write:
\begin{eqnarray}
\delta\hat{ \Psi}(x)=\int_0^\infty d\omega \sum_{I={\rm u-in,d1-in}} [ u_{I, \omega}(x) \hat{\gamma}_{I, \omega}+v^*_{I, \omega}(x) \hat{\gamma}^{\dag}_{I, \omega} ] \\
\nonumber
+\int_0^{\omega_{\rm max}} d\omega [ u_{\rm{d2-in, \omega}}(x) \hat{\gamma}^{\dag}_{\rm{d2-in, \omega}}+v^*_{\rm{d2-in, \omega}}(x) \hat{\gamma}_{\rm{d2-in, \omega}} ] \\ \; .
\end{eqnarray}
The normalization chosen in Eqs. \ref{d2-in.eq}-\ref{psiu.eq} guarantees that modes are normalized to unit quasiparticle current and so $[\hat{\gamma}_{I, \omega},\hat{\gamma}^{\dag}_{I', \omega'}]=\delta_{II'} \delta(\omega-\omega')$. An identical expression may be written changing in $\rightarrow$ out. Standard scattering theory arguments show that the $S(\omega)$-matrix coefficients connect the in-modes to the out-modes:
\begin{equation}\label{S-matrix.eq}
(\hat{\gamma}^{\dag}_{\rm{u-out, \omega}}, \hat{\gamma}^{\dag}_{\rm{d1-out, \omega}}, \hat{\gamma}_{\rm{d2-out, \omega}})=(\hat{\gamma}^{\dag}_{\rm{u-in, \omega}}, \hat{\gamma}^{\dag}_{\rm{d1-in, \omega}}, \hat{\gamma}_{\rm{d2-in, \omega}}) S^{\dag}(\omega) \; .
\end{equation}
Due to the pseudo-Hermitian character of the BdG equations (\ref{BdGDim.eq}),
this $S(\omega)$-matrix obeys, for frequencies $\omega<\omega_{\rm max}$, a pseudo-unitary relation $S^{\dag}(\omega)\eta S(\omega)=\eta$ with $\eta={\rm diag}(1,1,-1)$. This enforces non-standard relations among the transmission and reflection coefficients, for example, $|S_{\rm{d2d2}}(\omega)|^2-|S_{\rm{ud2}}(\omega)|^2-|S_{\rm{d1d2}}(\omega)|^2=|S_{\rm{ud1}}(\omega)|^2+|S_{\rm{d1d1}}
(\omega)|^2-|S_{\rm{d2d1}}(\omega)|^2=1$ (see Refs. \cite{zapata2009,recati2009,macher2009}).

The case $\omega>\omega_{\rm max}$ is simpler because the $S(\omega)$ matrix is unitary and there are no anomalous reflections or transmissions. Thus all asymptotic states can be chosen with $\nu>0$. If we rewrite $R_{I}(\omega):=S_{II}(\omega), T_{IJ}(\omega):=S_{IJ}(\omega)$ with $I,J$ different and taking values $I,J = u,d1$, then the usual unitarity relation $|R_{I}(\omega)|^2+|T_{IJ}(\omega)|^2=1$ applies.

The upstream phonon flux spectrum can be computed from these considerations and for $\omega<\omega_{\rm max}$ shows the remarkable form \cite{recati2009,macher2009}:
\begin{equation}\label{HawkingFlux.eq}
\frac{dI_{\rm{u-out}}}{d\omega}=|S_{\rm{uu}}(\omega)|^2 \frac{dI_{\rm{u-in}}}{d\omega}+|S_{\rm{ud1}}(\omega)|^2 \frac{dI_{\rm{d1-in}}}{d\omega}+|S_{\rm{ud2}}(\omega)|^2 \left(\frac{dI_{\rm{u-in}}}{d\omega}+1\right).
\end{equation}
Assuming that we can populate the ingoing fluxes with comoving thermal populations, namely, $dI_{\rm{\alpha-in}}/d\omega=n_B(\Omega_{\rm{\alpha-in}})$, where $\alpha=u,d1,d2$, $n_B(\Omega)=(e^{\beta\hbar\Omega}-1)^{-1}$ is the Bose-Einstein occupation at the common temperature $\beta^{-1}:=k_B T$, and $\Omega_{\alpha-\rm{in}}$ is the comoving frequency of the mode $\alpha-\rm{in}$:
\begin{equation}\label{Doppler.eq}
\Omega_{\rm{\alpha-in}}(\omega)=\omega-v_{\alpha} k_{\rm{\alpha-in}}(\omega),
\end{equation}
where $k_{\rm{\alpha-in}}(\omega)$ is the solution to Eq. (\ref{BogDR.eq}) for given $\omega$, with $v_{\alpha}=v_u,v_d$ the flow velocities. Equation (\ref{HawkingFlux.eq}) reveals that, even at $T=0$, a nonzero upstream flow of energy (phononic Hawking radiation) must be expected. The spectrum for $\omega>\omega_{\rm max}$ is of the same form as Eq. (\ref{HawkingFlux.eq}) but with the last term removed, i.e. without any zero-point energy flux.

\section{Hawking radiation spectra}

\begin{figure}[htb!]
\includegraphics[width=\columnwidth]{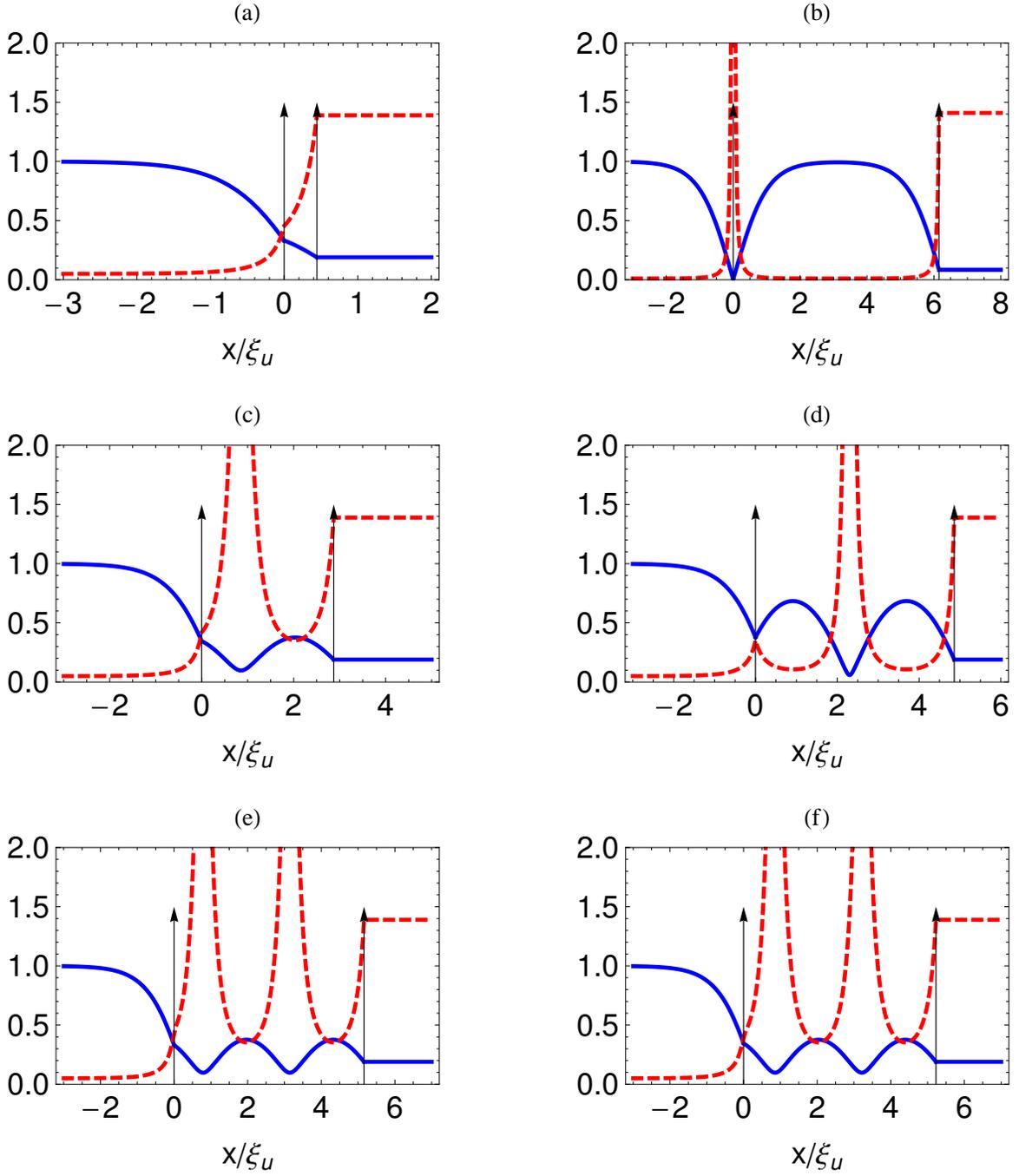}
\caption{As in Fig. \ref{figDDSamplePlot}, plot of the local speed of sound (solid blue), proportional to $\sqrt{\rho(x)}$ and the local speed of flow (dashed red), proportional to $1/\rho(x)$, for a variety of setups. These correspond to relevant Hawking radiation profiles to be shown in the next Fig. \ref{grHawkingPlot}. The parameters are for (a)-(f): number of added oscillations $0, 0, 1, 1, 2, 2$;  current $q\xi_u=0.05, 0.01, 0.05, 0.05, 0.05, 0.05$; delta barrier strength $z=0.989, 5.851, 0.989, 2.012, 0.989, 0.989$; of the two solutions for $\rho_0$ in Eq. (\ref{match2DeltasRho.eq}), that with the smallest $\rho_0$ is chosen in (a),(b),(e). The inter-barrier distance is $d/\xi_u=0.445, 6.150, 2.871, 4.861, 5.161, 5.229.$}
\label{grDdPlot}
\end{figure}

\begin{figure}[htb!]
\includegraphics[width=\columnwidth]{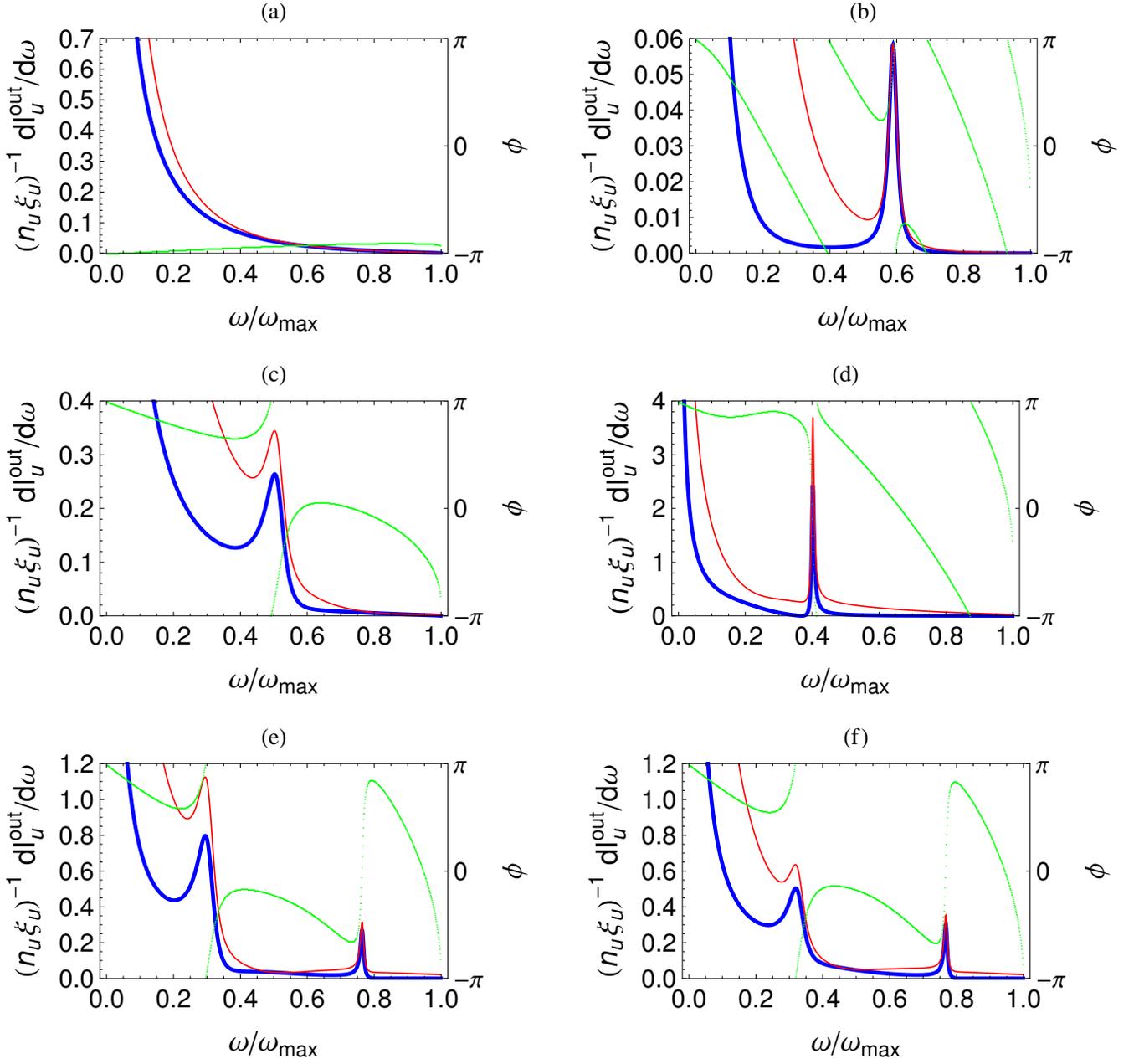}
\caption{Plot of an upstream spectral radiation profile [see Eq. (\ref{HawkingFlux.eq})], so called Hawking radiation. The parameters of the various graphs are identical to their counterparts in the previous figure. Thick blue: current spectrum at $k_B T=0$. Thin red: finite temperature (in units of $\mu$), 0.1 for the two uppermost curves and 0.3 for the rest). Dotted green: phase of the determinant of the $S(\omega)$ matrix. $\hbar\omega_{\rm max}/\mu=0.9299$ for all the graphs except (b), which has $\hbar\omega_{\rm max}/\mu=0.9859$. We draw the reader's attention to the fact that the peak in panel (d) corresponds to a dynamical instability, as can be seen from the drop of the green line as it approaches the peak for lower values of $\omega$, and not to a resonance, as it is the case in the other panels.}
\label{grHawkingPlot}
\end{figure}

Figure \ref{grHawkingPlot} shows some frequency spectra of the upstream phononic flow [see Eq. (\ref{HawkingFlux.eq})] for the setups and condensate solutions depicted previously in Fig. \ref{grDdPlot}, with a one-to-one correspondence between the graphs in the two figures. The top-left graph of Fig. \ref{grHawkingPlot} shows a structureless profile with a peak at $\omega=0$ followed by a $1/\omega$ tail, which is typical of Hawking radiation profiles in the absence of barriers, and very similar to what is obtained in the single delta barrier case. The message is that two nearby barriers behave similarly to a single barrier of double strength.
We may also note that the zero temperature contribution (thick blue line) is not easily distinguishable from the total contribution at nonzero temperature (thin red line), because at low frequencies they both follow the same thermal law $1/\omega$. This property seems to be common to all structures which do not permit one or more nonlinear oscillations of the condensate between the barriers. An exception to this trend occurs when the delta barrier strength $z$ is close to its upper limit. Then a bigger separation among barriers is possible and, as a consequence, a peak at nonzero $\omega$  may develop (see the upper-right graph at Fig. \ref{grHawkingPlot}). A general trend that can be clearly appreciated in the rest of the graphs is that the largest the separation the more peaks appear. In particular the two bottom graphs in Fig. \ref{grHawkingPlot} exhibit a double peak in the allowed frequency interval. The reader might wonder why panels (e) and (f) of Fig. \ref{grHawkingPlot} look somewhat different since the parameters hardly vary, the only difference being a relative change in the inter-barrier distance of approximately $0.01$. The reason is that the net amplification factor results from a rather complicated expression involving interferences between the scattering coefficients on both sides of the scattering region. We refer to Eq. (69) and Fig. 5, right panel of Ref. \cite{finazzi2010}, where a similar sensitive dependence has been found.

We note that spontaneous phonons are generated at the event horizons and, after multiple scattering by barriers and horizons, they are partly emitted into the subsonic side. Those scattering events include normal processes ($u$ and $v$ components do not mix) and Andreev (or anomalous) processes ($u$ converts into $v$ or viceversa, see. Ref. \cite{zapata2009}). The bigger the separation between the barriers, the more quasi-bound states can be accommodated between them and the more peaks appear in the Hawking radiation spectrum.

An important point which will be discussed in greater depth in the next section, is that peaks can be due to resonances (also called quasinormal modes) or instabilities, so more information is needed to classify them. An unequivocal experimental signature of Hawking radiation (associated to zero-point quantum fluctuations of an otherwise stationary state) would undoubtedly be favored by a transport regime where the classical flow is dynamically stable. We recall that the $S(\omega)$ matrix is pseudo-unitary in the Hawking sector $0<\omega<\omega_{\rm max}$. The general spectral theorem on these matrices \cite{mostafazadeh2004} guarantees that eigenvalues come in pairs of $s_i, 1/s_i^*$ or have unit modulus, $|s_j|=1$. A general trend we have observed is that in the region $\omega\alt\omega_{\rm max}/3$ most of scattering matrices show only one unit-modulus eigenvalue. By contrast, in the region $\omega\agt\omega_{\rm max}/2$, most of the $S$-matrices have three unimodular eigenvalues. Nevertheless, the theorem mentioned before guarantees that the determinant will be a pure phase. Then, as in a conventional phase shift analysis of quantum mechanical scattering, a phase jumping upwards when crossing through a peak from low to high $\omega$ can be interpreted as a resonance. On the contrary, a jump downwards will be reveal an instability. Hence, a most convenient way to detect resonances while distinguishing them from instabilities is to simultaneously plot the phase of the determinant of the $S(\omega)$ matrix. This corresponds to the thin green line appearing in all the graphs of  Fig. \ref{grHawkingPlot}. All curves show a clear resonance behavior, with the exception of the middle-right phase curve, which reveals an instability, as indicated by the sudden drop of the green line when traversing the instability. Most of the cases which we have studied show QNM behavior, but the the occurrence of instabilities is not so rare that it can be ignored. From our analysis, one cannot rule out the existence of strong instabilities with a large imaginary part in the eigenfrequency, since these would generate not easily detectable structures in the frequency spectrum. However, from the systematic analysis of Refs. \cite{finazzi2010, coutant2010}, where such instabilities where not found, one can conjecture that also in the present case those strong instabilities will not be found. An analysis of the time-dependent Gross-Pitaevskii equation in real time could establish that this is indeed the case. These considerations underline the need for a systematic search of instabilities and QNMs as poles of the propagator in the complex energy plane \cite{zapataTBP}.

\begin{figure}[htb!]
\includegraphics[width=\columnwidth]{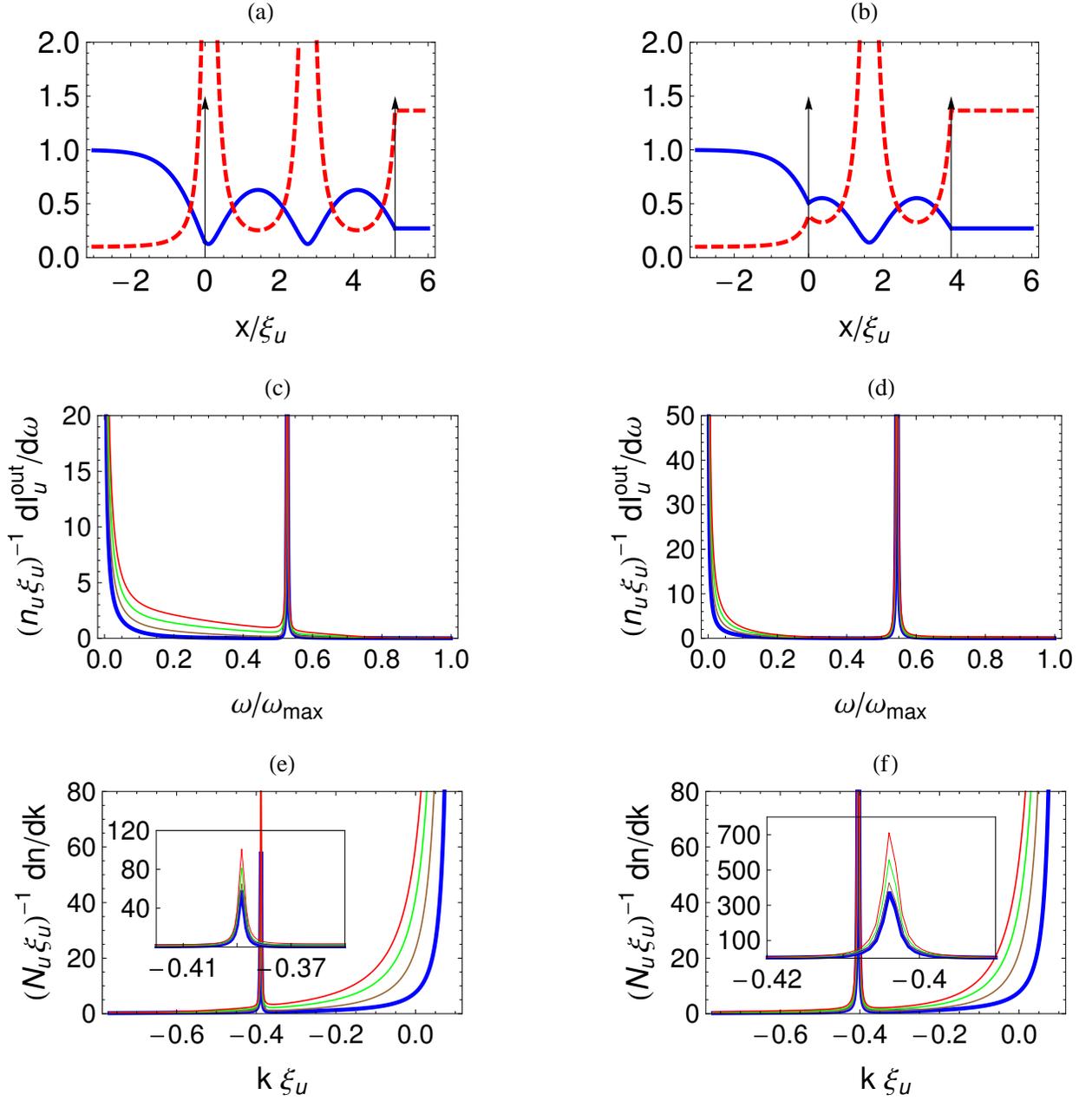}
\caption{Two setups of potential interest for a time-of-flight (TOF) experiment. Graphs in first/second row read as Figs.  \ref{grDdPlot}/\ref{grHawkingPlot} but with different parameters. The third row shows the  momentum spectral resolution on the left (subsonic) region. The negative momenta correspond to atoms traveling towards the left. Sharp peaks are mostly due to resonant Hawking radiation. The background tail reflects atoms in the depletion cloud moving to the left. All graphs share the parameters $q\xi_u=0.1$, $\hbar\omega_{\rm max}/\mu=0.8612$, one added oscillation, and of the two solutions for $\rho_0$ in Eq. (\ref{match2DeltasRho.eq}), that with the largest value is chosen. They differ  however in the barrier strength: z = 1.146, 0.968 for left and right column, respectively. In the second and third rows, the sequence of colors blue, brown, green, red, correspond to temperatures $k_B T /  \mu=0, 0.3, 0.6,0.9$, respectively. The insets in the third row show the span of the peaks in greater detail. The determinant of the $S(\omega)$ matrix is not shown but reveals that the peaks are due to resonances.}
\label{grTOFPlot}
\end{figure}

The examples  shown in Figs. \ref{grDdPlot}-\ref{grHawkingPlot} were chosen because they clearly reveal general trends. Unfortunately, in a time of flight (TOF) expansion, which correlates the long time density distribution with the momentum distribution of the initial state, their peaks barely stand out above the background of the depletion cloud, and even less when thermal fluctuations are taken into account. We have included in Fig. \ref{grTOFPlot} a set of two more favorable setups which show large signals for zero-point Hawking radiation. The two bottom graphs show the momentum distribution of the initial trapped state and are computed in the approximation where only the subsonic flow is included and boundary effects are neglected.
When a TOF measurement is performed in such systems, the depletion contribution is negligible at momentum values which however reveal clear resonant peaks. Zero-point Hawking radiation nearly exhausts those peaks at low temperatures and still gives the main contribution to the area under the peak at  temperatures as high as $0.9\mu$. This fact could allow for the unequivocal detection of Hawking radiation.

\section{Quasi-normal modes (resonances) and dynamical instabilities}\label{QNM.sect}

A discussion of instabilities in BEC black-hole analogs can be found in Refs. \cite{leonhardt2003,finazzi2010,leonhardt2003A,barcelo2006}. For a study of QNMs in a similar context, we refer to Ref. \cite{barcelo2007}. General studies of QNM in gravitational black holes and in optical analogs can be found in Refs. \cite{kokkotas1999} and \cite{ching1998,settimi2009}, respectively. A more complete study for the setup considered in this article will be presented \cite{zapataTBP}.

We have said that peaks in the spectrum of Hawking radiation may be due to resonances or instabilities. The former are characterized by poles of the analytical continuation of the $S(\omega)$ matrix in the complex $\omega$-plane with Im$(\omega)<0$ while the latter have Im$(\omega)>0$. Moreover, both types of complex modes may have $|\rm{Im}(\omega)|$ so large that they are not clearly appreciated in the radiation spectrum, yet they can hide important instabilities. While a systematic search for poles is left for a future work, we discuss here some general features of the behavior of QNMs and instabilities in the complex $k$-plane.

When a linear stability analysis is made of a stationary solution of the GP equation (\ref{GPDim.eq}), instabilities appear as solutions of the BdG equations (\ref{BdGDim.eq}) whose frequency $\omega$ has a positive imaginary part. Because complex frequency implies complex asymptotic wave numbers, these solutions must be localized in real space. For a given complex frequency, there are always four complex $k$'s (counting each different $k$ with its multiplicity), two with Im$(k)>0$ and two with Im$(k)<0$.

Like in the search for bound states in conventional quantum mechanics, a discrete, possibly empty set of modes is to be expected. The mode with the largest Im$(\omega)$ dominates the long time behavior and its inverse is a good measure of the decay time of the condensate due to that instability. There is a well defined procedure to quantize such unstable modes \cite{rossignoli2005}, which show up as essentially free particles (instead of oscillators). In open systems like the presently analyzed, there are however some subtleties in that quantization procedure which, to our knowledge, have not been properly addressed in the literature. Specifically, some spectral properties of the BdG operators are used which can only be guaranteed for finite-dimensional operators \cite{brink2001}.

\begin{figure}[htb!]
\includegraphics[width=\columnwidth]{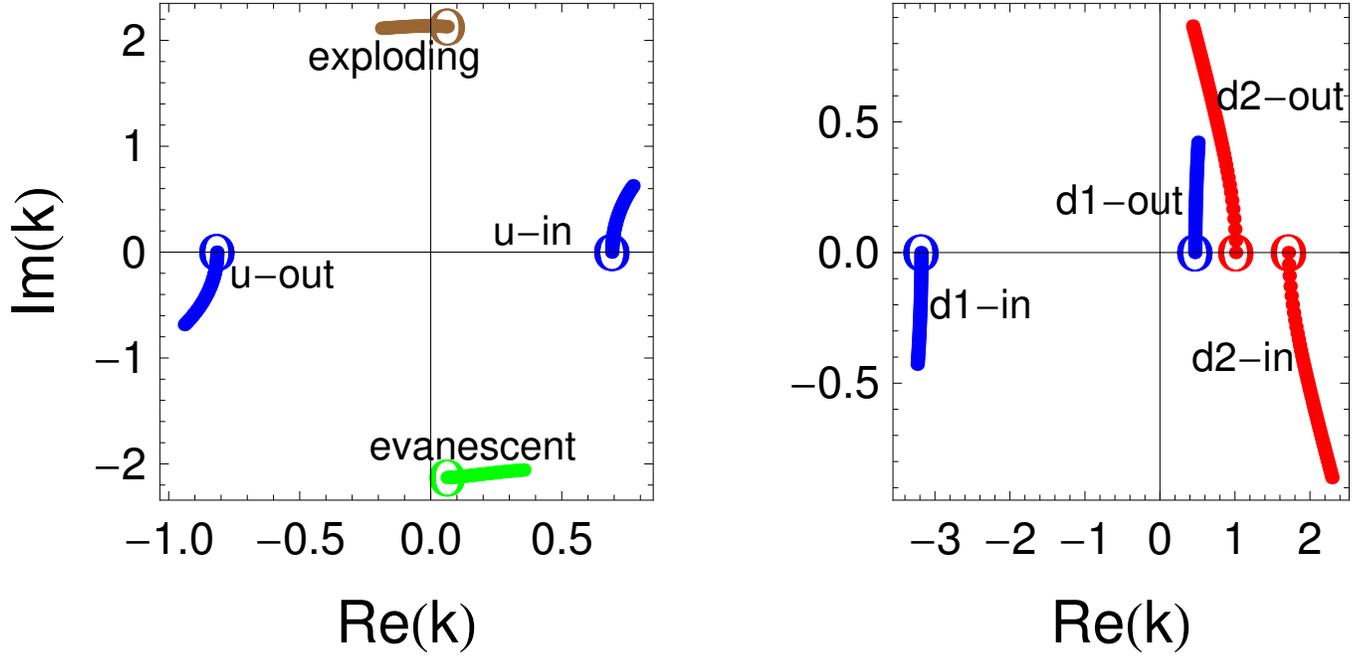}
\caption{Migration in the complex $k$-plane of the roots of the Bogoliubov dispersion relation Eq. (\ref{BogDR.eq}), for the two asymptotic sides (left is subsonic, righ is supersonic) of a BH configuration. Units of $\xi_u^{-1}$ (left) or $\xi_d^{-1}$ (right) are used. The paths shown result from tracking the $k$ solutions as Im$(\omega)$ evolves from zero to a positive value, leaving the real part fixed. The labels and colors correspond to those of Fig. \ref{figDispRelation} but we have included the evanescent (green) and  exploding (brown) solutions. The parameters are chosen so that the structure has a current $q=0.01$ and Re$(\omega)$ is fixed to a value $0.8<\omega_{\rm max}$.}
\label{figComplexKPlot}
\end{figure}

At a given real frequency between 0 and $\omega_{\rm max}$ we may have have four complex $k$ solutions on each side of the interface. These are shown by circles in Fig. \ref{figComplexKPlot}. On the upstream side (left graph), they correspond to an incoming, outgoing, evanescent, and exploding solution. Downstream (right graph) we have two incoming and two outgoing solutions, corresponding to the normal (d1) and anomalous (d2) channels. For instance, a conventional retarded scattering state is characterized by one incoming channel matching all possible outgoing and evanescent solutions, with no amplitude for exploding or other incoming solutions.

An instability is a bound state made exclusively of spatially decaying solutions that happen to match at a particular complex frequency (with positive imaginary part). Localized solutions correspond to Im$(k)<0$ on the upstream side and to Im$(k)>0$ on the downstream side. Figure \ref{figComplexKPlot} shows how the various $k$ solutions evolve as $\omega$ varies from Im$(\omega)=0$ to Im$(\omega)>0$ while Re$(\omega)$ remains constant. Eventual instabilities can only be obtained by matching the evolved ``u-out'' and ``evanescent'' solutions upstream (Im$k<0$) and the evolved ``d1-out" and ``d2-out" downstream (Im$k>0$), with vanishing amplitudes for the other solutions. We emphasize that the wave function of these unstable modes involves exclusively analytical continuations of outgoing and evanescent solutions.

QNM modes, on the other hand, are obtained from the analytical continuation to the Im$(\omega)<0$ half-plane of the retarded Green's function of the time-dependent version of the Bogoliubov-deGennes equations (\ref{BdGDim.eq}) (see Refs. \cite{ching1998,settimi2009} for a discussion in an optical context but precisely translatable to the present one). More specifically, these QNMs can be obtained as an analytical continuation to Im$(\omega)<0$ of scattering states involving only outgoing and evanescent solutions, i.e. without the intervention of incoming waves. Like for instabilities, a discrete number of QNMs is to be expected. Therefore, in order to find resonances, we are only interested in the evolution of ``u-out'' and ``evanescent'' solutions on the upstream side, and ``d1-out" and ``d2-out" on the downstream side, ruling out the other solutions. Interestingly, this is exactly the same set of solutions which are relevant in the search for instabilities. We conclude that instabilities and resonances are obtained from the matching of the same set of solutions albeit in different regions of the complex energy plane.

Hence, QNM/instabilities correspond to poles of the $S(\omega)$ matrix when analytically continued to the lower/upper complex plane from the real energy line. This can be readily seen if one allows for a general coefficient in the in-wave component of the scattering solution Eq. (\ref{d2-in.eq}), for example. So both resonances and instabilities are candidates to explain peaks in the square of a given $S(\omega)$ matrix coefficient. The frequency of those possible underlying modes must of course have an imaginary part much smaller than their real part. Since the determinant of the $S(\omega)$ matrix is a pure phase (as noted in the previous section), it can be used to discriminate between both behaviors. Specifically, a jump upwards/downwards of the phase when traversing the peak with increasing frequency, implies a QNM/instability.

Figure \ref{grHawkingPlot}, together with other not shown data, reveals that peaks in the Hawking spectrum are mostly due to resonances, instabilities being more the exception than the norm. While this sampling is encouraging, a systematic search for QNMs and instabilities in various stationary flows is left for future work.

Finally we note that, both for QNMs and instabilities, the boundary conditions here described are different from those adopted in Ref. \cite{barcelo2006,barcelo2007}.

\section{Discussion and conclusion}

We have studied the flow of an atomic condensate through a double barrier interface separating regions of subsonic and supersonic flow. Such a setup provides a scenario where Hawking radiation into the subsonic side is enhanced at some frequencies due to multiple scattering of quasiparticles by the two barriers and the modulations of the condensate. The resulting highly non-thermal Hawking radiation presents peaks at frequencies which may lie well above the working temperature and thus can be unambiguously interpreted as stemming from quantum fluctuations of the quasiparticle vacuum. The non-thermal Hawking spectrum emitted by the double barrier interface represents and important advantage over the cases of single or zero barrier, where the low-frequency zero-point radiation has a thermal character which makes it more difficult to distinguish from genuinely thermal radiation.

Our calculation has been based on a model of stationary flow of condensate and quasiparticles through a double barrier structure with open boundary conditions, where the condensate density is asymptotically flat on both sides of the structure, while quasiparticle motion is described by scattering states characterized by incoming channels that are thermally populated. While a stationary scattering picture of transport has proved to hold predictive power in electron systems, it still represents an idealized scenario in cold atom contexts, where stationary circuits have not yet been developed and where transport of finite-sized condensates is mostly investigated within a time-dependent scheme \cite{Aspect2009}. As long as steady condensate transport is still an item for the future, it will be of interest to perform numerical simulations of time-dependent transport that may reveal those features of the Andreev-Hawking phenomena which we have explored here.

An important question is whether, in currently achievable setups, the resonance frequency can be tuned to lie well above the currently attainable temperatures that would characterize the incoming quasiparticle population. We have seen that, while in a time-of-flight experiment the contribution from Hawking radiation peaks may be easily overshadowed by the contribution of the depletion cloud, setups can be designed where the resonant Hawking peaks are sufficiently sharp to be clearly visible even at temperatures comparable to the chemical potential.

We thank A. Aspect, C. D\'{\i}az Guerra, L. Garay, P. Leboeuf, N. Pavloff, G. V. Shlyapnikov, and C. Westbrook for valuable discussions. This work has been supported by the joint France-Spain Acci\'on Integrada HF2008-0088 (PHC - Picasso Program). Support from MICINN (Spain) through grants FIS2007-65723 and FIS2010-21372, from Comunidad de Madrid through grant MICROSERES-CM (S2009/TIC-1476), and from the Swiss National Science Foundation, is also acknowledged.

\appendix

\section{Mean-field calculation.}\label{MFT.app}

In this and the next appendices we will use units such that $\hbar=m=\xi_u=1$, where $\xi_{u}:=\hbar/\sqrt{mg_{\rm 1D}n_u}$ is the asymptotic coherence length on the subsonic (upstream). This means in particular that the unit of frequency and energy is $g_{\rm 1D}n_u$. We will also rescale the wave functions $\Psi,u,v$ by $1/\sqrt{n_u}$. Then the equations in these units can be immediately read from Eqs. (\ref{GPDim.eq})-(\ref{BdGDim.eq}) by making $\hbar=m=g_{\rm 1D}=1$, which amounts to taking $n_u\equiv \lim_{x\rightarrow -\infty}|\Psi_0(x)|^2=1$ and $c_u=1$. The time-independent Gross-Pitaevskki equation (\ref{GPDim.eq}) for the (scaled) condensate wave-function $\Psi_0(x)$ reads \cite{pethick2002,dalfovo1999,leggett2001}:
\begin{equation}\label{GP.eq}
\left[-\frac{1}{2}\frac{d^2}{dx^2}-\mu+V_{\rm ext}(x)+|\Psi_0(x)|^2\right]\Psi_0(x)=0
\end{equation}

We consider potential profiles $V_{\rm ext}(x)$ consisting of one or two delta barriers, namely $V_{1}(x):=z\delta(x)$ and $V_{2}(x):=z[\delta(x)+\delta(x-d)]$. The condensate will be asymptotically flat and subsonic to the left of the first barrier and flat and supersonic to the right of the last $\delta$ barrier. Then, we can choose phases such that for $x<0$ the condensate profile reads \cite{danshita2006}:

\begin{eqnarray}\label{sbCond.eq}
\nonumber
\Psi_0(x)&=&e^{i (q x+\theta_0)}\left[\gamma(x)+i q\right] \\
\nonumber
e^{i \theta_0}&:=&\frac{\gamma(0)-i q}{\sqrt{\gamma(0)^2+q^2}} \\
\gamma(x)&:=&\sqrt{1-q^2}\tanh\left[\sqrt{1-q^2}(x_0-x)\right] \; .
\end{eqnarray}
We note that $x_0\geqslant0$, which means that $\rho'(0^-)<0$, where $\rho(x):=|\Psi_0(x)|^2\rightarrow 1$ as $x\rightarrow -\infty$, see below. The chemical potential and uniform current read

\begin{eqnarray}\label{chemPotCurrent.eq}
\nonumber
\mu&=&1+\frac{q^2}{2} \\
j&=&q \; ,
\end{eqnarray}
where  $0\leqslant q<1$ is required to have $j>0$ and subsonic flow.

On the supersonic side, the condensate wave function is a simple plane wave, which by current conservation has to be of the form:

\begin{eqnarray}\label{spCond.eq}
\nonumber
\Psi_0(x)&=&A_{\rm min} e^{i \left(q x/A^2_{\rm min}+\phi_a\right)} \\
A_{\rm min}&:=&\frac{\sqrt{q^2+q\sqrt{8+q^2}}}{2}<1\; .
\end{eqnarray}
where $\phi_a$ is a phase to be determined later.

\begin{figure}[htb!]
\includegraphics[width=\columnwidth]{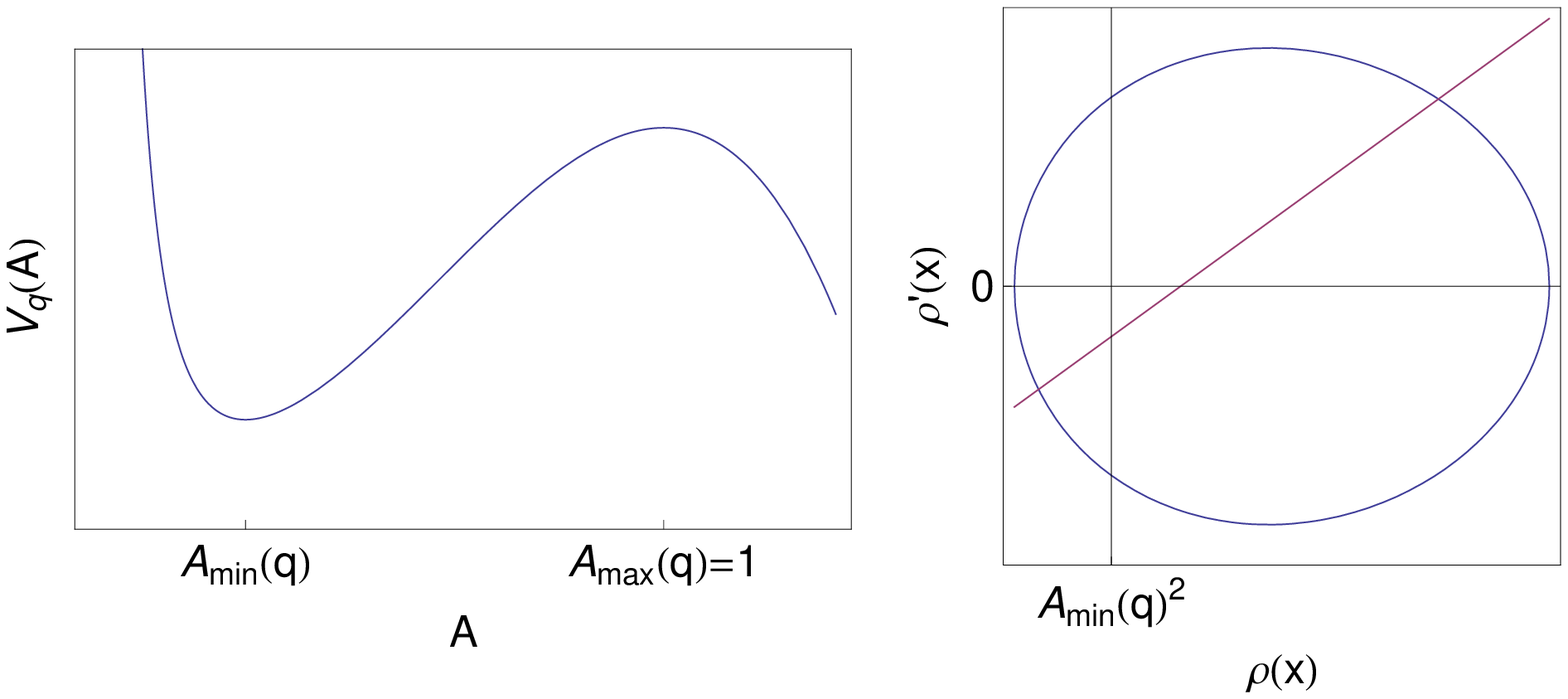}
\caption{Left graph: plot of the effective potential experienced by the fictitious particle presented in the discussion of Eq. (\ref{pot.Equiv.eq}), for a typical value of $q$ between 0 and 1. Right graph: plot of Eqs. (\ref{match2DeltasRho.eq}) (in blue the first, in red the second) for a typical case with an intermediate value of $z$.}
\label{figMFTVarPlot}
\end{figure}

There is a simple analogy which permits to qualitatively understand the behavior of the condensate wave-function. Shifting to an amplitude and phase representation, and splitting the GP equation (\ref{GP.eq}) into its real and imaginary components in the regions where $V_{\rm ext}(x)=0$, we may write

\begin{eqnarray}\label{GP.ampphase.eq}
\nonumber
\Psi_0(x)&=&A(x)e^{i \phi(x)} \\
\nonumber
\frac{A''(x)}{2}&+&\mu A(x) -A(x)^3-\frac{j^2}{2A(x)^3}=0 \\
\frac{d}{dx}j&=&\frac{d}{dx}\left[A(x)^2\phi'(x)\right]=0\; .
\end{eqnarray}
The third equation, which comes from the imaginary part of Eq. (\ref{GP.eq}), is but the continuity equation. The second equation, after multiplication by $A'(x)$ and one integration, can be written as [cf. Eq. (\ref{chemPotCurrent.eq})]:

\begin{eqnarray}\label{pot.Equiv.eq}
\nonumber
E&=&\frac{A'(x)^2}{2}+V_q\left(A(x)\right) \\
V_q(A)&:=&\left(1+\frac{q^2}{2}\right)A^2+\frac{q^2}{2A^2}-\frac{A^4}{2} \; ,
\end{eqnarray}
$E$ being an integration constant. Therefore (see e.g. Ref. \cite{langer1966,zapata1996}) by looking at the position as a time coordinate and the amplitude as the position of a fictitious particle, this equation expresses the energy conservation for this particle under a force deriving from the potential $V_q(A)$ (see left Fig. \ref{figMFTVarPlot}). For $q<1$ and $A>0$ this potential has a minimum at $A_{\rm min}$ [see Eq. (\ref{spCond.eq})], on the left of a maximum at $A_{\rm max}=1$. The flat solutions [$A'(x)=0$] at the minimum/maximum are supersonic/subsonic [see Eq. (\ref{spCond.eq}) and Ref. \cite{leboeuf2001}]. Integration of this equation with the initial condition $\lim_{x\rightarrow-\infty}A(x)= 1$ leads to (\ref{sbCond.eq}). We note that the integration constant must be chosen as $E_{\rm max}\equiv V_q(A_{\rm max})=V_q(1)$. We will also use the quantity $E_{\rm min}\equiv V_q(A_{\rm min})$. A delta barrier $z\delta(x-x_0)$ appears as an instantaneous kick, occurring at ``time" $x_0$, which adds a positive amount to the speed of the particle, which now becomes $A'(x_0^+)=A'(x_0^-)+2 z A(x_0)$. The ``energy'' of the particle changes instantaneously without changing the phase $\phi(x_0)$ or ``position" $A(x_0)$. In both the single- and double-barrier case, the first kick, $\delta(x)$, must take place when the particle has negative ``speed'', $A'(0^-)<0$, otherwise the kick would increase the energy with no chance of ending at $A_{\rm min}$. The case of a single delta barrier at the origin, $V_{1}(x)$, leads to a simple change in energy which can be solved to obtain $z(q)$ [or $q(z)$, as the dependence is monotonic, see Fig. \ref{figDeltaStrengthPlot}]:

\begin{eqnarray}\label{1deltaMatch.eq}
\nonumber
E_{\rm max}-E_{\rm min}&=&2 z^2 A_{\rm min}^2 \\
z(q)&=&\sqrt{\frac{E_{\rm max}-E_{\rm min}}{2 A^2_{\rm min}}} \; ,
\end{eqnarray}
and the parameter $x_0$ can be extracted from (\ref{sbCond.eq}) by imposing $|\Psi_0(x_0)|=A_{\rm min}$.

The case of two delta barriers, $V_{2}(x)$, is more involved. In this case, the energy after the first kick must lie between the two limiting values $E_{\rm min}<E<E_{\rm max}$, because if bigger than $E_{\rm max}$, it would mean that the particle has acquired a positive speed, and the second kick could never send it to $A_{\rm min}$. To end up in $A_{\rm min}$ with zero kinetic energy, $A'(d^-)<0$ is required. We conclude that, for a given current $q$, the maximum strength $z$ which allows for a solution is (\ref{1deltaMatch.eq}), which is the unique $z$ value for single delta-barrier. From the matching at $x=d$ it follows that the energy for the particle motion between kicks ($0<x<d$) is

\begin{equation}\label{energyBetweenDeltas.eq}
\frac{A'(x)^2}{2}+V_q[A(x)]=E_{\rm min}+2 z^2 A_{\rm min}=:E(q,z) \; .
\end{equation}
Analysis of the $x=0$ kick leads to another equation which, combined with (\ref{energyBetweenDeltas.eq}) computed at $x=0^+$, leads to the following two equations:

\begin{equation}\label{match2Deltas.eq}
E(q,z) = \frac{A'(0^+)^2}{2} + V_q\left[A(0)\right] = \frac{1}{2} + q^2 + 2 z A(0) A'(0^+)-2 z^2 A(0)^2
\end{equation}
which allows us to solve for $A(0), A'(0^+)$ as a function of $q,z$. If we introduce the densities $\rho(x)=A^2(x), \rho_0:=\rho(0), \rho_0':=\rho'(0^+)$, then Eq. (\ref{match2Deltas.eq}) can be rewritten:

\begin{eqnarray}\label{match2DeltasRho.eq}
\nonumber
\frac{\rho_0'^2}{8}&=&\rho_0\left[E(q,z)-V_q\left(\sqrt{\rho_0}\right)\right] \\
\rho_0'&=&2 z \rho_0 -\frac{q^2+1/2-E(q,z)}{z} \; ,
\end{eqnarray}
whose plot is shown in right Fig. \ref{figMFTVarPlot} \cite{footnote2}. The minimum $z$ which allows for a solution can be computed by enforcing that the two lines cross tangentially (see blue line in Fig. \ref{figDeltaStrengthPlot}).

From the previous discussion,  a method can be outlined to compute the possible distances $d$ between the barriers for given $q,z$ between the relevant limits: One takes one of the two solutions of (\ref{match2DeltasRho.eq}), and propagates it (clockwise in the blue line of right Fig. \ref{figMFTVarPlot}) until $\rho(x)=A^2_{\rm min}, \rho'(x)<0$. The ``time" $x$ needed is the minimum distance between barriers. An integer number of periods can be added, where the period is the time needed to get back to the initial point. Two families of solutions result, corresponding to the two solutions that can be chosen from Eq. (\ref{match2DeltasRho.eq}).

To obtain explicit formulae, it is necessary to solve for the motion between the delta barriers. A straightforward method is to integrate Eq. (\ref{energyBetweenDeltas.eq}), written in the density representation:

\begin{eqnarray}\label{rhoMotionBetweenDeltas.eq}
\nonumber
\frac{\rho'(x)^2}{4}&=&\left(\rho(x)-1\right)^2\left(\rho(x)-q^2\right)-2 \Delta \! E(q,z) \rho(x) \\
\Delta \! E(q,z)&:=&E_{\rm max}-E(q,z) \geq 0 \; ,
\end{eqnarray}
the last inequality requiring $z<z(q)$, which will be assumed hereafter. The polynomial in $\rho(x)$ on the r.h.s. of the first equation has three real roots ordered such that $q^2<e_1<e_2<1<e_3$. The integral is tabulated (see Ref. \cite{byrd1971}) and the solution is:

\begin{eqnarray}\label{rhoBetweenDeltas.eq}
\nonumber
\rho(x)&=&e_1+(e_2-e_1) \textrm{sn}^2\left(\sqrt{e_3-e_1}x-F_0, k\right) \\
\nonumber
F_0&:=&\left\{\begin{array}[c]{ll}
              F\left(\arcsin\left(\sqrt{\frac{\rho_0-e_1}{e_2-e_1}}\right),k\right), & \rho_0'>0 \\
              2 K(k)- F\left(\arcsin\left(\sqrt{\frac{\rho_0-e_1}{e_2-e_1}}\right),k\right), & \rho_0'<0
            \end{array}
\right. \\
\nonumber
k&:=&\sqrt{\frac{e_2-e_1}{e_3-e_1}},
\end{eqnarray}
where $F(\phi,k):=\int_0^\phi d\theta/\sqrt{1-k^2 \sin^2(\theta)},K(k):=F(\pi/2,k)$ are the incomplete and complete elliptic integrals of the first kind, and $\textrm{sn}(u,k)$ is a Jacobi elliptic function, whose period is $4 K(k)$. Some straightforward algebra allows to compute the possible inter-barrier distances in explicit form:

\begin{eqnarray}\label{ddPossibleDistances.eq}
\nonumber
d&=&\frac{\alpha-F_0+2 n K(k)}{\sqrt{e_3-e_1}}, \; n \in \mathbb{Z} \; (d>0) \\
\nonumber
\alpha&:=&\left\{
\begin{array}{cc}
  \alpha_1, & \textrm{cn}(\alpha1,k)<0 \\
  2K(k)-\alpha_1, & \textrm{cn}(\alpha1,k)>0
\end{array}
\right. \\
\alpha_1&:=&\textrm{sn}^{-1}\left(\sqrt{\frac{A_{\rm min}^2-e_1}{e_2-e_1}},k\right).
\end{eqnarray}
This identifies $2K(k)/\sqrt{e_3-e_1}$ as the "period", i.e., the distance that can be added to the deltas to have another solution in the same family, as mentioned in the paragraph following the paragraph of Eq. (\ref{match2DeltasRho.eq}). A plot of the possible $d$ solutions for four possible values of the current is shown in Fig. \ref{figPhaseDiagplot}.

Finally, the phase $\phi(x)$ can be computed from the last Eq. (\ref{GP.ampphase.eq}):

\begin{eqnarray}\label{phiBetweenDeltas.eq}
\nonumber
\phi(x)-\phi(0)&=&q \int_0^x \frac{dx'}{\rho(x')} \\
&=&\frac{q}{e_1\sqrt{e_3-e_1}}\left\{\Pi\left(\textrm{am}(\sqrt{e_3-e_1}x+F_0,k),1-\frac{e_2}{e_1},k\right)
-\Pi\left(\textrm{am}(F_0,k),1-\frac{e_2}{e_1},k\right)\right\}
\end{eqnarray}
where, here and in the previous equations, $\textrm{cn}(u,k),\textrm{am}(u,k)$ are Jacobi elliptic functions, and $\Pi(\phi,m,k):=\int_0^\phi d\theta \{(1-m \sin^2\theta)\sqrt{1-k^2 \sin^2\theta}\}^{-1}$ is the incomplete elliptic integral of the third kind.

We summarize below the algorithms to compute the complete GP solution:

{\it Single delta barrier algorithm.}
Start with a given current $0<q<1$. Get the unique delta barrier strength, $z$, from Eq. (\ref{1deltaMatch.eq}). Compute $x_0>0$ from Eq. (\ref{sbCond.eq}). The solution is then Eq. (\ref{sbCond.eq}) for $x<0$ and Eq. (\ref{spCond.eq}) with $\phi_a=0$ for $x>0$.

{\it Double delta barrier algorithm.}
Start with a given current $0<q<1$. Compute the minimum delta barrier strength, $z_{\rm min}(q)$, as that which makes the two equations (\ref{match2DeltasRho.eq}) tangent in the $\rho_0, \rho_0'$ plane. Compute the maximum delta barrier strength, $z_{\rm max}(q)=z(q)$ from Eq. (\ref{1deltaMatch.eq}). Choose $z_{\rm min}(q)<z<z_{\rm max}(q)$ and solve for $\rho_0, \rho_0'$ in Eq. (\ref{match2DeltasRho.eq}) . Choose one of the two solutions. Find the roots of the polynomial Eq. (\ref{rhoMotionBetweenDeltas.eq}) and order them $e_1<e_2<e_3$. The expression for $\rho(x)$ between deltas can be obtained from Eq. (\ref{rhoBetweenDeltas.eq}). Take one of the possible distances, $d$, from Eq. (\ref{ddPossibleDistances.eq}). Compute $x_0>0$ from Eq. (\ref{sbCond.eq}) and the chosen value of $\rho_0$. The solution is Eq. (\ref{sbCond.eq}) for $x<0$, Eqs. (\ref{rhoBetweenDeltas.eq}), (\ref{phiBetweenDeltas.eq}) with $\phi(0)=0$ for $0<x<d$, and Eq. (\ref{spCond.eq}) with $\phi_a=\phi(d)$ from Eq. (\ref{phiBetweenDeltas.eq}) for $x>d$.

\section{Exact solution of B${\bf d}$G equations.}\label{BdG.app}

\subsection{Subsonic region}

For a general reference on the content of this Appendix, see Refs. \cite{faddeev2007, chen1998}. There are some sign conventions in those references which are not followed in the present work. It can be shown that a solution of the time-dependent GP equation (with no external potential),

\begin{equation}\label{TDGP.eq}
i \frac{\partial \Psi(x,t)}{\partial t } = -\frac{1}{2} \frac{\partial^2 \Psi(x,t)}{\partial x^2}+|\Psi(x,t)|^2 \Psi(x,t) \; ,
\end{equation}
makes the following system over-determined:

\begin{eqnarray}\label{ZS.eq}
\nonumber
\frac{\partial}{\partial x} \left[
                              \begin{array}{c}
                                w_1(x,t) \\
                                w_2(x,t) \\
                              \end{array}
                            \right] &=&
                            \left[
                              \begin{array}{cc}
                                -i \lambda & \Psi^*(x,t) \\
                                \Psi(x,t) & i\lambda \\
                              \end{array}
                            \right]
                            \left[
                              \begin{array}{c}
                                w_1(x,t) \\
                                w_2(x,t) \\
                              \end{array}
                            \right] \\
\frac{\partial}{\partial t} \left[
                              \begin{array}{c}
                                w_1(x,t) \\
                                w_2(x,t) \\
                              \end{array}
                            \right] &=&
                            \left[
                              \begin{array}{cc}
                                i |\Psi(x,t)|^2/2+i\lambda^2 & -i \partial_x \Psi^*(x,t) /2- \lambda \Psi^*(x,t) \\
                                i \partial_x \Psi(x,t) /2- \lambda \Psi(x,t) & -i |\Psi(x,t)|^2/2-i\lambda^2 \\
                              \end{array}
                            \right]
                            \left[
                              \begin{array}{c}
                                w_1(x,t) \\
                                w_2(x,t) \\
                              \end{array}
                            \right] \; .
\end{eqnarray}
This is the so-called Lax pair of equations of the Zakharov-Shabat problem. The spinor solutions with certain boundary conditions of the first equation are denominated Jost functions. Here $\lambda$ is an auxiliary scattering parameter which allows to map the general solution of the Cauchy initial value problem posed by the GP equation into a linear scattering problem at $t=0$ from $\Psi(x,0)$. The time evolution of the scattering data is easily computed, and from the results at time $t$ on can reconstruct $\Psi(x,t)$ using inverse scattering techniques. Then, it can be easily shown that the squared Jost functions solve the linear perturbation problem associated to Eq. (\ref{TDGP.eq}), which is described by the time-dependent BdG equations (see Refs. \cite{pethick2002,dalfovo1999,leggett2001}):

\begin{equation}\label{TDBdG.eq}
i\partial_t
\left[
  \begin{array}{c}
    w_2^2(x,t) \\
    w_1^2(x,t) \\
  \end{array}
\right]
=
\left[
  \begin{array}{cc}
    -\partial_{xx}/2 + 2|\Psi(x,t)|^2 & \Psi(x,t)^2 \\
    -\Psi^*(x,t)^2 & \partial_{xx}/2 - 2|\Psi(x,t)|^2) \\
  \end{array}
\right]
\left[
  \begin{array}{c}
    w_2^2(x,t) \\
    w_1^2(x,t) \\
  \end{array}
\right] \; ,
\end{equation}
where we note the non-obvious order of the spinor components.

In the particular case case of a stationary, asymptotically flat, and subsonic condensate solution, $\Psi(x,t)=e^{-i \mu t}\Psi_0(x)$, where $\Psi_0(x)$ is given in Eq. (\ref{sbCond.eq}) [see also Eq. (\ref{chemPotCurrent.eq})], a general stationary solution of (\ref{ZS.eq}) is:

\begin{equation} \label{ansatz-w2-w1.eq}
\left[
  \begin{array}{c}
    w_1(x,t) \\
    w_2(x,t) \\
  \end{array}
\right]= e^{i(k x-\varepsilon t)/2}
\left[
  \begin{array}{c}
    e^{i(\mu t-q x -\theta_0)/2} v_1(x) \\
    e^{-i(\mu t-q x -\theta_0)/2} v_2(x) \\
  \end{array}
\right]
\end{equation}
where $v'_i(-\infty)=0$ ($i=1,2$). The variables $k/2, \varepsilon/2$ are introduced in such a way that $k$ and $\varepsilon$ can be identified with the momentum and energy of the excitation after squaring. The phases are chosen to cancel the asymptotic phase of the condensate (except for a phase $\sqrt{1-q^2}+i q$)  [see Eq. (\ref{sbCond.eq})]. Substituting ansatz (\ref{ansatz-w2-w1.eq}) into the first Eq. (\ref{ZS.eq}) leads to:

\begin{equation}\label{ZS.v.eq}
\frac{\partial}{\partial x} \left[
                              \begin{array}{c}
                                v_1(x) \\
                                v_2(x) \\
                              \end{array}
                            \right] =
                            \left[
                              \begin{array}{cc}
                                -i [\lambda+(k-q)/2] & \gamma(x)-iq \\
                                \gamma(x)+iq & i [\lambda-(k+q)/2] \\
                              \end{array}
                            \right]
                            \left[
                              \begin{array}{c}
                                v_1(x) \\
                                v_2(x) \\
                              \end{array}
                            \right]
\end{equation}
which in the limit $x\rightarrow -\infty $ and using the $v'_i(-\infty)=0$ boundary condition, yields $\lambda$:

\begin{equation}\label{lambda.eq}
\lambda = \frac{q}{2}\pm \sqrt{\frac{k^2}{4}+1}\; .
\end{equation}

Equation (\ref{ZS.v.eq}) can be solved by adding and subtracting both equations, which introduces $S(x):=v_1(x)+v_2(x), T(x):=v_1(x)-v_2(x)$. Solving for $S(x)$ as a function of $T(x), T'(x)$, and introducing the result into the other equation, leads to $T''(x)+i k T'(x)=0$, whose only solution satisfying the boundary condition $v'_i(-\infty)=0$ [which implies $T'(-\infty)=0$] is a constant $T(x)=T_0$. Some further algebra leads to:
\begin{eqnarray}
\nonumber
v_1(x)&=&\frac{S(x)+T(x)}{2}=\frac{T_0}{2}\left[\frac{i\gamma(x)-k/2}{q/2+\lambda}+1\right]\\
v_2(x)&=&\frac{S(x)-T(x)}{2}=\frac{T_0}{2}\left[\frac{i\gamma(x)-k/2}{q/2+\lambda}-1\right].
\end{eqnarray}

To solve for the spectrum, $\varepsilon$, we use the $\partial_t$ equation of (\ref{ZS.eq}). Due to the over-determinacy of Eqs. (\ref{ZS.eq}) when $\Psi(x,t)$ is a solution of Eq. (\ref{TDGP.eq}), we only need one of them, e.g. the first one. Solving this equation in the limit $x\rightarrow -\infty$, the Bogoliubov spectrum is found to be [after invoking Eqs. (\ref{lambda.eq}) and (\ref{chemPotCurrent.eq})]:
\begin{equation}\label{BogSbSP.eq}
\varepsilon = q k \pm k \sqrt{1+\frac{k^2}{4}},
\end{equation}
where $\pm$ correlates with the same symbol in Eq. (\ref{lambda.eq}). From Eqs. (\ref{lambda.eq}),(\ref{BogSbSP.eq}) we get $\lambda+q/2=\varepsilon/k$, and demanding that $|u(-\infty,t)|^2-|v(-\infty,t)|^2=1$, we can write the modes as:

\begin{equation}\label{BdGSbModes.eq}
\left[
  \begin{array}{c}
    u(x,t) \\
    v(x,t) \\
  \end{array}
\right] =
e^{i(k x-\varepsilon t)}\sqrt{\frac{\varepsilon}{8 k^2(\varepsilon-k q)}}
\left[
  \begin{array}{c}
    e^{i(q x -\mu t +\theta_0)}\left(1+\frac{k}{2\varepsilon}(k-2i\gamma(x))\right)^2 \\
    e^{-i(q x -\mu t +\theta_0)}\left(1-\frac{k}{2\varepsilon}(k-2i\gamma(x))\right)^2 \\
  \end{array}
\right].
\end{equation}

\subsection{Supersonic region.}

In this case there is no need to use the technique of squaring the Jost functions, as the solution can be trivially worked out. We only quote the results, using the notation of Eq. (\ref{spCond.eq}) and introducing $p:=q/A_{\rm min}, c:=A_{\rm min}$, which are, respectively, the condensate wave vector and the speed of sound in the supersonic region:
\begin{eqnarray}\label{BdGSsModes.eq}
\nonumber
\left[
  \begin{array}{c}
    u(x,t) \\
    v(x,t) \\
  \end{array}
\right] &=&
e^{i(k x-\varepsilon t)}
\left[
  \begin{array}{c}
    e^{i(p x -\mu t +\phi_a)} G_{+}(k,q)\\
    -e^{-i(p x -\mu t +\phi_a)} G_{-}(k,q)\\
  \end{array}
\right] \\
\left[
  \begin{array}{c}
    u(x,t) \\
    v(x,t) \\
  \end{array}
\right] &=&
e^{i(k x-\varepsilon t)}
\left[
  \begin{array}{c}
    -e^{i(p x -\mu t +\phi_a)} G_{-}(k,q)\\
    e^{-i(p x -\mu t +\phi_a)} G_{+}(k,q)\\
  \end{array}
\right].
\end{eqnarray}
where $G_{\pm}(k,q)$ is a short-hand notation for
\begin{equation}
G_{\pm}(k,q)\equiv \left[\frac{k^2/2+c^2}{\sqrt{k^2(k^2+4c^2)}}\pm\frac{1}{2}\right]^{1/2}
\end{equation}
We are assuming, without loss of generality, that $\varepsilon>0$. The first solution(\ref{BdGSsModes.eq}) is always valid and is normalized to $|u(x,t)|^2-|v(x,t)|^2=1$. The second solution is propagating only when $\varepsilon<\omega_{\rm max}$ and its normalization is anomalous: $|u(x,t)|^2-|v(x,t)|^2=-1$. As expected, the spectrum is that of Bogoliubov quasiparticles for the supersonic region:

\begin{equation}\label{BogSpSP.eq}
\varepsilon = p k \pm \sqrt{\frac{k^2}{4}\left(2c^2+\frac{k^2}{2}\right)}.
\end{equation}

\bigskip

\begin{thebibliography}{99}
\bibitem{hawking1974} S. W. Hawking, Nature {\bf 248}, 30 (1974); Commun. Math. Phys. {\bf 43}, 199 (1975).
\bibitem{Unruh1976} W. G. Unruh, Phys. Rev. D {\bf 14}, 870 (1976).
\bibitem{Unruh1981} W. G. Unruh, Phys. Rev. Lett. {\bf 46}, 1351 (1981).
\bibitem{Garay2000} L. J. Garay, J. R. Anglin, J. I. Cirac, P. Zoller, Phys. Rev. Lett. {\bf 85}, 4643 (2000); Phys. Rev. A {\bf 63}, 023611 (2001).
\bibitem{Laughlin2003} G. Chapline, R. B. Laughlin and D. I. Santiago, in {\it Analog Models of General Relativity}, M. Visser ed. (World Scientific, Singapore, 2003).
\bibitem{balbinot2008} R. Balbinot, A. Fabri, S. Fagnocchi, A. Recati and I. Carusotto, Phys. Rev. A {\bf 78}, 021603 (2008).
\bibitem{carusotto2008} I. Carusotto, S. Fagnocchi, A. Recati, R. Balbinot and A. Fabri, New J. Phys. {\bf 10}, 103001 (2008).
\bibitem{macher2009} J. Macher and R. Parentani, Phys. Rev. A {\bf 80}, 043601 (2009).
\bibitem{finazzi2010} S. Finazzi and R. Parentani, New J. Phys. {\bf 12}, 095015 (2010).
\bibitem{coutant2010} A. Coutant and R. Parentani, Phys. Rev. D {\bf 81}, 084042 (2010).
\bibitem{zapata2009} I. Zapata and F. Sols, Phys. Rev. Lett. {\bf 102}, 180405 (2009).
\bibitem{leonhardt2008} U. Leonhardt and T.G. Philbin, in {\it Quantum Analogues: From Phase Transitions to Black Holes and Cosmology}, W. G. Unruh and R. Schutzhold eds. (Springer, Berlin, 2007); arXiv:0803.0669.
\bibitem{corley1999} S. Corley and T. Jacobson, Phys. Rev. D {\bf59}, 124011 (1999).
\bibitem{recati2009} A. Recati, N. Pavloff and I. Carusotto, Phys. Rev. A {\bf 80}, 043603 (2009).
\bibitem{lahav2010} O. Lahav, A. Ital, A. Blumkin, C. Gordon, S. Rinott, A. Zayats and J. Steinhauer, Phys. Rev. Lett. {\bf 105}, 240401 (2010).
\bibitem{Unruh1995} W. G. Unruh, Phys. Rev. D {\bf 51}, 2827 (1995).
\bibitem{finazzi2011} S. Finazzi and R. Parentani, arXiv:1012.1556, to appear in Phys. Rev. D.
\bibitem{jackson1998} A.D. Jackson, G.M. Kavoulakis, and C.J. Pethick, Phys. Rev. A 58, 2417 (1998).
\bibitem{leboeuf2001} P. Leboeuf and N. Pavloff, Phys. Rev. A \textbf{64}, 033602 (2001).
\bibitem{pavloffTBP} N. Pavloff and I. Carusotto, to be published.
\bibitem{barcelo2006a} C. Barcel\'o, S. Liberati, S. Sonego and M. Visser, Phys. Rev. Lett. {\bf 97}, 171301 (2006).
\bibitem{mayoral2011} C. Mayoral, A. Recati, A. Fabbri, R. Parentani, R. Balbinot, and I. Carusotto, New J. Phys. {\bf 13} 024007 (2011).
\bibitem{birrell1982} N. D. Birrell, P. C. W. Davies, {\it Quantum Fields in Curved Space} (Cambridge University Press, Cambridge, 1982).
\bibitem{leonhardt2003} U. Leonhardt, T. Kiss, and P. \"{O}hberg, J. Opt. B {\bf 5}, S42 (2003).
\bibitem{leonhardt2003A} U. Leonhardt, T. Kiss, and P. \"{O}hberg, Phys. Rev. A {\bf 67}, 033602 (2003).
\bibitem{unruh2005} W. G. Unruh and R. Sch\"{u}tzhold, Phys. Rev. D {\bf 71}, 024028 (2005).
\bibitem{chen1998} X.-J. CHen, Z.-D. CHen and N.-N. Huang, J. Phys. A: Math. Gen. {\bf 31}, 6929 (1998).
\bibitem{zapataTBP} I. Zapata {\it et al.}, to be published.
\bibitem{footnote1} This reflects the well-known symmetry that for each solution with frequency $\omega$ and wave function $(u,v)$, there exists another (physically identical) solution with frequency $-\omega^*$, wave function $(v^*,u^*)$, and opposite normalization.
\bibitem{mostafazadeh2004} A. Mostrafazadeh, J. Math. Phys. {\bf 45}, 932 (2004).
\bibitem{barcelo2006} C. Barcel\'o, A. Cano, L. J. Garay and G. Jannes, Phys. Rev. D {\bf 74}, 024008 (2006).
\bibitem{barcelo2007} C. Barcel\'o, A. Cano, L. J. Garay and G. Jannes, Phys. Rev. D {\bf 75}, 084024 (2007).
\bibitem{kokkotas1999} K. D. Kokkotas and B. G. Schmidt, Living Rev. Relativity {\bf 2}, 2 (1999), http://www.livingreviews.org/lrr-1999-2.
\bibitem{ching1998} E. S. C. Ching, P. T. Leung, A. Maassen van den Brink, W. M. Suen, S. S. Tong and K. Young, Rev. Mod. Phys. {\bf 70}, 1545 (1998).
\bibitem{settimi2009} A. Settimi, S. Severini and B. J. Hoenders, J. Opt. Soc. Am. B {\bf 26}, 876 (2009).
\bibitem{rossignoli2005} R. Rossignoli and A. M. Kowalski, Phys. Rev. A {\bf 72}, 032101 (2005).
\bibitem{brink2001} A. M. van den Brink and K. Young, J. Phys. A: Math. Gen. {\bf 34}, 2607 (2001).
\bibitem{Aspect2009} A. Aspect and M. Inguscio, Phys. Today \textbf{62}, 30 (2009).
\bibitem{pethick2002} C. J. Pethick and H. Smith, {\it Bose-Einstein Condensation in Dilute Gases} (Cambridge University Press, Cambridge, 2002).
\bibitem{dalfovo1999} F. Dalfovo \textit{et al.}, Rev. Mod. Phys. \textbf{71}, 463 (1999).
\bibitem{leggett2001} A.~J. Leggett, Rev. Mod. Phys. \textbf{73}, 307 (2001).
\bibitem{danshita2006} I. Danshita, N. Yokoshi and S. Kurihara, New Journal of Physics {\bf 8}, 44 (2006).
\bibitem{langer1966} J. S. Langer, V. Ambegaokar, Phys. Rev. {\bf 164}, 498 (1967) .
\bibitem{zapata1996} I. Zapata and F. Sols, Phys. Rev. B {\bf 53}, 6693 (1996).
\bibitem{footnote2} There are other solutions involving branches of the first Eq. (\ref{match2DeltasRho.eq}) with values of $\rho_0$ greater than 1 which thus must be discarded as unphysical.
\bibitem{byrd1971} P. F. Byrd and M. D. Friedman, {\it Handbook of Elliptic Integrals for Engineers and Scientists} (Springer-Verlag, Berlin, 1971).
\bibitem{faddeev2007} L. D. Faddeev and L. A. Takhtajan, {\it Hamiltoninan Methods in the Theory of Solitons} (Springer-Verlag, Berlin, 2007).
\end{thebibliography}
\end{document}